\documentclass[twocolumn]{aastex62}

\usepackage{graphicx}
\usepackage{subfigure}
\usepackage{amsmath}  
\usepackage{natbib}  

\received{10-Nov-2021}
\revised{23-Feb-2022}
\accepted{25-Apr-2022}
\submitjournal{ApJ}

\shorttitle{A spectroscopic study of blue supergiant stars in M 31 and M 33}
\shortauthors{Liu et al.}

\begin{document}

\title{A spectroscopic study of blue supergiant stars in Local Group spiral galaxies: Andromeda and Triangulum
}


\correspondingauthor{Cheng Liu}
\email{cheng@nao.cas.cn}

\author{Cheng Liu}
\affil{Beijing Planetarium, Beijing Academy of Science and Technology, Beijing, 100044, China}

\author{Rolf-Peter Kudritzki}
\affiliation{Institute for Astronomy, University of Hawaii, 2680 Woodlawn Drive, Honolulu, HI 96822, USA}
\affiliation{Universit\"atssternwarte, Ludwig-Maximilian-Universit\"at M\"unchen, Scheinerstr. 1, 81679 M\"unchen, Germany}

\author{Gang Zhao}
\affiliation{CAS Key Laboratory of Optical Astronomy, National Astronomical Observatories, Chinese Academy of Sciences, Beijing 100101, China}
\affiliation{School of Astronomy and Space Science, University of Chinese Academy of Sciences, Beijing 100049, China}

\author{Miguel A. Urbaneja}
\affiliation{Institut f\"{u} Astro- und Teilchenphysik, Universit\"{a}t Innsbruck, Technikerstr. 25/8, A-6020 Innsbruck, Austria}

\author{Yang Huang}
\affiliation{South-Western Institute for Astronomy Research, Yunnan University, Kunming 650500, China}

\author{Huawei Zhang}
\affiliation{Department of Astronomy, School of Physics, Peking University, Beijing 100871, China}
\affiliation{Kavli Institute for Astronomy and Astrophysics, Peking University, Beijing 100871, China}

\author{Jingkun Zhao}
\affiliation{CAS Key Laboratory of Optical Astronomy, National Astronomical Observatories, Chinese Academy of Sciences, Beijing 100101, China}



\begin{abstract}

Low-resolution LAMOST and Keck spectra of blue supergiant stars distributed over the disks of the Local Group spiral galaxies M 31 and M 33 are analyzed to determine stellar effective temperatures, gravities, metallicities, and reddening. Logarithmic metallicities at the center of the galaxies (in solar units) of $0.30\pm0.09$ and $0.11\pm0.04$ and  metallicity gradients of $-0.37\pm0.13$ dex/$R_{25}$ and $-0.36\pm0.16$ dex/$R_{25}$ are measured for M 31 and M 33, respectively. For M 33 the 2-dimensional distribution of metallicity indicates a deviation from azimutal symmetry with an off-centre peak. The flux-weighted gravity-luminosity relationship of blue supergiant stars is used to determine a distance modulus of 24.51$\pm$0.13 mag for M 31 and  24.93$\pm$0.07 mag for M 33. For M 31 the flux-weighted gravity--luminosity relationship (FGLR) distance agrees well with other methods. For M 33 the FGLR-based distance is larger than the distances from Cepheids studies but it is in good agreement with work on eclipsing binaries, planetary nebulae , long-period variables, and the tip of the red giant branch.

\end{abstract}


\section{Introduction} \label{sec:intro}
Measuring chemical composition and distances of galaxies is crucial for our understanding  of galaxy formation and evolution and for constraining the cosmological parameters. Distances provide the basis for the determination of the Hubble constant and are essential to characterize galaxy mass, radii and luminosities, whereas the distribution of chemical composition of stars and the interstellar medium accross the galaxies reflects the signature of the complex dynamics of galaxy evolution, such as, galaxy merging, gas infall, galactic winds, star formation history, and initial mass function and serves as a valuable constraint for detailed chemical evolution modeling \citep{1999PASP..111..919H, 2009ApJ...696..729M, 2015MNRAS.450..342K}.

In the seminal work by \cite{1942ApJ....95...52A}, galactic metallicity gradients - the linear decrease of the logarithm of heavy element abundances with galactocentric radius - have been studied for the first time. Extensive work following this pioneering study has then revealed that in the local Universe practically all spiral galaxies show a radial decrease in metallicity \citep{2014A&A...563A..49S, 2017MNRAS.469..151B}. Usually emission lines from H {\scriptsize $\mathrm{II}$} regions are used to examine metallicity gradients of spiral galaxies \citep{1999AJ....118.2775G, 2012ApJ...758..133S, 2016MNRAS.458.1866T} and the galaxy mass-metallicity relation, the relationship between average galactic metallicity and stellar mass \citep{1979A&A....80..155L, 2004ApJ...613..898T}. Metallicities, mostly restricted to oxygen (O) abundances, are obtained in most cases through the analysis of a few very strong emission lines. However, the metallicities determined with these strong-line methods strongly depend on the calibration used and are subject to large systematic uncertainties. For example, \cite{2008ApJ...681.1183K} have found that the absolute metallicity scale of galaxies varies up to 0.7 dex that the slope of the mass-metallicity relationship of galaxies can change greatly.  \cite{2009ApJ...700..309B} in their study of galaxy NGC 300, demonstrate that absolute values of metallicity can shift by as much as 0.6 dex and that metallicity gradients can change significantly when using the different calibrations of the strong-line method. In contrast to the strong line methods, the additional use of weak auroral emission lines, the so-called direct method, is much more accurate, however, this technique requires long exposure times. In addition, as also has become apparent more recently, from the work using the weak auroral lines, derived abundances might be subject to systematic uncertainties due to oxygen depletion onto dust grains and a possible detection bias toward lower abundances at high metallicities \citep{2012MNRAS.427.1463Z, 2016ApJ...830...64B}.

An alternative to the use of H {\scriptsize $\mathrm{II}$} regions is the quantitative absorption spectroscopy of individual blue supergiant stars (BSG). BSGs of B and A spectral types are massive stars in the stellar mass range between 12 and 60 $M_{\odot}$  in a short post main-sequence evolutionary stage. The absolute visual magnitudes of BSGs can reach up to $M_{\rm{V}} \approx -9.5$ mag, rivaling the integrated light of globular clusters and dwarf galaxies and allowing for spectroscopy of these objects in galaxies out to 10 Mpc distance  with current instruments. In contrast to abundance determinations using H {\scriptsize $\mathrm{II}$} regions, which only provide light elements (e.g., He, N and O) abundances, the quantitative high-resolution spectral analysis of BSGs can additionally provide metallicity from iron group elemenmts elements \citep[see][]{2000ApJ...541..610V, 2006A&A...445.1099P, 2008A&A...479..849S}. With the development of an efficient new spectral analysis technique and using stellar atmosphere model spectra calculated with detailed NLTE atomic models, \cite{2008ApJ...681..269K} have demonstrated for the Sculptor group galaxy NGC 300 that low-resolution spectra (FWHM $\sim$ 5 \AA) with sufficient signal-to-noise ratio (SNR)  allow for an accurate determination of stellar parameters, interstellar reddening, extinction and metallicity. Since then, this work has been applied to a variety of Local Group galaxies (WLM,  \citealt{2008ApJ...684..118U}, M33, \citealt{2009ApJ...704.1120U} , NGC 3109, \citealt{2014ApJ...785..151H}, IC 1613, \citealt{2018ApJ...860..130B}) and beyond (M81, \citealt{2012ApJ...747...15K} , NGC4248, \citealt{2013ApJ...779L..20K}, NGC 3621, \citealt{2014ApJ...788...56K} ,NGC 55, \citealt{2016ApJ...829...70K}, M83, \citealt{2016ApJ...830...64B}). Besides the investigation of only a few objects in M 31 \citep{2000ApJ...541..610V, 2001MNRAS.325..257S, 2002A&A...395..519T, 2006astro.ph.11044P}, no comprehensive spectroscopic study of BSG has been carried out in M 31 so far.

In this paper, we present the spectral analysis of low-resolution LAMOST spectra of BSGs in the disk of the spiral galaxies M 31 and M 33. Additional Keck DEIMOS spectra of BSGs in M 33 disk are also analyzed in order to enlarge our stellar sample. Both galaxies belong to the Local Group and are ideal laboratory to measure metallicity gradient in disk and to investigate how gradients evolve with time. In addition, we use the BSGs of our samples as accurate distance indicators through the FGLR, which has been introduced by \cite{2003ApJ...582L..83K, 2008ApJ...681..269K}. The concept of the FGLR is based on the assumption that massive stars evolve through the B and A supergiant stage at roughly constant luminosity and mass. This leads to a tight correlation of flux-weighted gravity (a combination of gravity and temperature) with absolute bolometric magnitude. Since reddening is a by-product of the spectral analysis, the determined distance via FGLR is free of the uncertainties caused by interstellar reddening. The FGLR method has been applied to determine the distances for several galaxies and it was found that FGLR-based distances are consistent with distances determined by other methods, demonstrating the reliability of the method \citep{2008ApJ...684..118U, 2012ApJ...747...15K, 2013ApJ...779L..20K, 2014ApJ...788...56K, 2014ApJ...785..151H}.

In Section \ref{sec:sample}, we describe the observations of target stars and the selections of BSG candidates which belong to M 31 and M 33, respectively. We determine extinction, effective temperature, gravity, and metallicity through the quantitative spectral analysis method in Section \ref{sec:analysis} and explain the reasons  why we can not obtain stellar parameters for close half of BSG sample. Section \ref
{sec:red-sp} discusses interstellar reddening and stellar properties by comparing the spectroscopically determined stellar parameters with evolutionary tracks. In Section \ref{sec:MG}, we measure the metallicity gradients of BSGs for both M 31 and M 33, respectively, and compare our results with the published trends of metallicities constrained from the spectra of H {\scriptsize $\mathrm{II}$} regions and PNe. Using the FGLR method in Section \ref{sect:dist}, we determine a new distance to M 31 and M 33 from BSG samples, and compare the new results with distances determined from other methods. Finally, we present our conclusions in Section \ref{sec:con}.

\section{Target Stars and Observations} \label{sec:sample}
The spectra of supergiant stars in the direction of our M 31 and M 33 fields were obtained as part of the LAMOST survey \citep{2012RAA....12.1197C, 2006ChJAA...6..265Z, 2012RAA....12..723Z}. LAMOST, also known as the Guo Shou Jing Telescope, is a new type of 5$^{o}$ wide field telescope with a large aperture of 4 m. Its focal plane is covered by 4000 fibers connected to 16 sets of multi-object optical spectrometers. Those spectrometers simultaneously obtain a similar number of low-resolution ($R \sim 1800$) spectra in two wavelength regions,  3700--5900 and 5700--9000 $\rm{\AA}$. Spectra are then wavelength calibrated and intensity normalized following the procedures in a 1D pipeline \citep{2012RAA....12.1243L}. In this work, radial velocity determined from pipeline LSP3 \citep{2015MNRAS.448..822X} are used for  M 31 and M 33 member selection. For spectra with a SNR greater than 10, the typical uncertainty of the radial velocity is $\sim$5 km/s and 10 to 15 km/s for a poorer spectrum of lower SNR.

For the  observations a catalog of supergiant star candidates has been submitted to LAMOST by Huang et al. (in prep.) for M 31 and M 33 fields and observations were carried out since 2010. In this work, we select BSG candidates from this catalog. To compile the catalog, targets were initially selected based on their colors from the contents of the Local Group Galaxy Survey \citep[LGGS,][]{2006AJ....131.2478M}. In order to distinguish genuine supergiants from foreground dwarfs, the same selection criteria as in \cite{2006AJ....131.2478M}, \cite{2009ApJ...703..441D} and \cite{2009ApJ...703..420M} for blue supergiants, yellow supergiants and red supergiants were used simultaneously. To provide an adequate SNR of the spectra, the supergiant candidates were selected to have $V <$ 19. Finally, to avoid light contamination, all nearby objects must be dimmer by 2 mag within 4 arcsec. We note that an independent spectroscopic survey of M 31 and M 33 for blue supergiant stars, luminous blue variables and emission line stars has been carried by Roberta Humphreys and collaborators \citep{2013ApJ...773...46H, 2014ApJ...790...48H, 2017ApJ...844...40H, 2017ApJ...836...64H, 2016ApJ...825...50G}. The spectra of this survey have been made available to us most recently and a quantitative spectroscopic investigation is planned as a follow-up to this work.

\subsection{Identifying BSGs in M 31} \label{m31selection}

863 supergiant candidates have been observed in M 31 fields in the first phase of LAMOST survey. Of these objects
565 BSG candidates have a Johnson $Q$-index \citep{2006AJ....131.2478M} of Q $\le$ 0.0 mag ensuring that the spectral types are most likely in a suitable range for our analysis. Following \cite{2009ApJ...703..441D} and using radial velocity measurements Huang et al. (in prep.) distinguished two \textquotedblleft ranks" of M 31 members in the catalog. Rank 1 stars are \textquotedblleft mostly certain" M31 members, while rank 2 stars are \textquotedblleft probable" M31 members. Most of excluded objects are the foreground dwarfs classified as rank 3. In addition to radial velocity, the absolute bolometric magnitude limit ($<$ --10.3 for M31 and $<$ --10.9 for M33) was used to remove contaminations from halo giants or binaries as discussed in Sect. \ref{Results}. Except for the main magnitude cut of V $<$ 19, the impact on the luminosity distribution of the sample could be ignored.
As most of stars are dimmer than 16 mag in $V$ magnitude, we restrict the SNR of the spectra to values higher than 10 in order to determine accurate stellar parameters for our targets. As a consequence, we obtain a list of 34 rank 1 and 33 rank 2 objects.

Some of the M 31 candidates in the catalogue are not stars, but small clusters. This was confirmed by examining images of the LGGS and HST carefully. We found that seven objects have sizes larger than nearby stars in the same image. They turned out as previously known clusters \citep{2007AA...471..127G}. 18 objects have been classified as foreground dwarfs by \cite{2016AJ....152...62M} and \cite{2016ApJ...825...50G}. One bright object (J004510.04+413657.6) classified as a foreground contaminant in the catalogue turned out to be confirmed M 31 A-type supergiant \citep{2000ApJ...541..610V} and is added to our list of BSGs for the forthcoming spectra analysis (see Table \ref{M31_M33_sample}). Recently a new luminous blue variables (LBVs) -- J003720.65+401637.7 was discovered by \cite{2019ApJ...884L...7H}. We find that the spectrum of the target does not show clear emission lines in the spectral ranes of our analysis (see below). Therefore, it will inclucded in spectral analysis in next section.

As most of objects have a relatively low signal-to-noise ratio, we decided to convolve all LAMOST spectra to have a resolution of full width at half-maximum (FWHM) of 5 \AA. In this way we obtain a reasonable signal at a resolution sufficient for our quantitative spectral analysis purpose \citep{2008ApJ...681..269K, 2012ApJ...747...15K, 2014ApJ...785..151H}. This convolution is also done for the LAMOST spectra of the M 33 fields. 

\subsection{Identifying BSGs in M 33}

\subsubsection{Stars observed with LAMOST}
According to the catalog, spectra of 537 supergiant candidates have been observed in M 33 fields in the first phase of LAMOST survey. The same methodology as described in Sect. \ref{m31selection} is used to identify BSG candidates for the spectral analysis. We obtain a list of 15 mostly certain and 34 probable members from the catalog. Nine objects were identified as foreground dwarfs in previous work by \cite{2012ApJ...750...97D} and \cite{2016AJ....152...62M}. One star (J013344.43+303843.9) is classified as a dwarf by our method, however, it was spectroscopically confirmed as a late B-type supergiant in M33 by \cite{2009ApJ...704.1120U}. It is, therefore,  included in our sample for the spectral analysis in the next section. 

\subsubsection{Stars observed with Keck DEIMOS}
To enlarge our sample, we also collect high SNR spectra of 30 BSG candidates (see Table \ref{M31_M33_sample}) which were observed with the Keck Telescope on Mauna Kea in 2003. The observations were taken with the DEIMOS spectrograph on 2003 November 1 with good seeing (0$''$.6) using a 1200 l/mm grating and yielding a dispersion of 0.33 \AA~pixel$^{-1}$ and a spectral resolution (FWHM) of 1.6 \AA. The spectra cover a range of 3700 to 9000 \AA. Among the 30 targets, three stars, IFM-B 600,1004, and 1026, have been classified as Luminous blue variables (LBVs) by \cite{1953ApJ...118..353H} and \cite{1996ApJ...469..629M}. We find that the spectrum of IFM-B 600 does not show clear emission lines in the spectral windows used for our analysis. Therefore, the spectrum of this star is also analyzed in the next section. Two stars are removed from our sample, because their spectra are dominated many emission lines. Although the Keck DEIMOS spectra have enough SNR for a spectral analysis, they are also convolved to obtain a final FWHM = 5 \AA. In this way, the Keck data set is compatible with the LAMOST data.

\startlongtable
\begin{deluxetable*}{rlrrrrrrrrccr}
\tablecaption{BSG candidates with the observation of LAMOST and Keck DEIMOS \label{M31_M33_sample}}
\tablewidth{0pt}
\tablehead{
\colhead{ID} & \colhead{$\alpha(J2000.0)$} & \colhead{$\delta(J2000.0)$} & \colhead{$R/R_{25}$} & \colhead{Sp.T.} & \colhead{$V$} & \colhead{$B-V$} &\colhead{$V_{our}$} &\colhead{$V_{\rm{our}}-V_{\rm{exp}}$} &\colhead{SNR} &\colhead{Rank} &\colhead{Ref.}\\
\colhead{} &\colhead{} & \colhead{} & \colhead{} & \colhead{} & \colhead{(mag)} & \colhead{(mag)} &\colhead{km s$^{-1}$} &\colhead{km s$^{-1}$} &\colhead{} &\colhead{} &\colhead{}
}
\startdata
J003720.65+401637.7 &9.3360417    &+40.2771389 &1.08 &LBV &15.854 &0.231 &--472 &26 &35 &1 &a, b\\
J003728.99+402007.8 &9.3707917    &+40.3355000 &1.07 &B1I &17.310 &0.107 &--471 &20 &19 &1 &a, c\\
J004005.02+403242.2 &10.0209167  &+40.5450556 &0.57 & &17.901 &0.055 &--509 &21 &15 &1 &a\\
J004021.21+403117.1 &10.0883750  &+40.5214167 &0.62 &F0I &16.648 &0.275 &--532 &--23 &20 &1 &a, d\\
J004033.90+403047.1 &10.1412500  &+40.5130833 &0.66 &BI &17.747 &0.111 &--577 &--87 &16 &1 &a, c\\
J004036.92+410119.0 &10.1538333  &+41.0219444 &0.56 & &17.933 &0.198 &--740 &--320 &10 &1 &a\\
J004051.59+403303.0 &10.2149583  &+40.5508333 &0.67 &LBV &16.989 &0.216 &--629 &--155 &27 &1 &a\\
J004126.23+405214.3  &10.3592917  &+40.8706389 &0.32 &A6I &17.485 &0.308 &--591 &--79 &20 &1 &a\\
J004133.62+404208.4  &10.3900833  &+40.7023333 &0.61 & &17.841 &0.021 &--479 &--38 &15 &1 &a\\
J004154.82+405706.7  &10.4784167  &+40.9518611 &0.30 &B8I &17.910 &0.168 &--484 &--12 &16 &1 &a\\
J004202.86+411434.6  &10.5119167  &+41.2429444 &0.26 & &18.088 &0.256 &--373 &--17 &20 &1 &a\\
J004212.27+413527.4  &10.5511250  &+41.5909444 &0.81 &B8I &17.541 &0.225 &--545 &--287 &11 &2 &a\\
J004253.42+412700.5  &10.7225833  &+41.4501389 &0.28 & &17.265 &0.012 &--401 &--196 &13 &2 &a\\
J004422.84+420433.1  &11.0951667  &+42.0758611 &0.90 &B0.5I &16.465 &0.017 &--387 &--233 &14 &2 &a, d\\
J004510.04+413657.6  &11.2918333  &+41.6160000 &        &A3aI &16.336 &0.260 &--116 &20 &17 &3 &a, e\\
J004535.26+413238.6  &11.3969167  &+41.5440556 &0.81 & &16.788 &0.600 &--160 &29 &16 &2 &a\\
J004623.14+413847.5  &11.5964167  &+41.6465278 &1.00 &B8Ie &16.140 &0.137 &--240 &--59 &22 &2 &a\\
\hline
J013242.51+302455.3  &23.1771250   &+30.4153611 &0.61 & &15.406 &0.258 &--114 &1 &20 &2 &a\\
J013300.23+302323.7 &23.2509583  &+30.3899167 &0.56 &A0Ia &16.440 &0.142 &--68 &36 &11 &2 &a\\
J013322.43+303513.2 &23.3434583  &+30.5870000 &0.23 & &18.363 &-0.064 &--109 &17 &15 &2 &a\\
J013323.57+302221.8 &23.3482083  &+30.3727222 &0.51 & &16.628 &0.600 &--58 &42 &30 &2 &a\\
J013324.14+300520.9  &23.3505833   &+30.0891389 &0.99 & &16.850 &0.610 &--182 &--76 &37 &2 &a\\
J013337.09+303521.6  &23.4045417   &+30.5893333 &0.15 &A2I &17.937 &0.062 &--130 &--25 &21 &2 &a, f\\
J013344.66+303631.6  &23.4360833   &+30.6087778 &0.09 &A6III &16.904 &0.031 &--200 &--100 &11 &2 &a\\
J013351.20+303224.5  &23.4633333   &+30.5401389 &0.21 &B9I &17.194 &0.034 &--153 &--36 &12 &2 &g\\
J013356.89+300900.6 &23.4870417  &+30.1501667 &0.91 & &17.789 &0.551 &--151 &--33 &21 &2 &a\\
J013359.74+304124.4 &23.4989167  &+30.6901111 &0.08 &B2I &17.424 &0.035 &--200 &48 &26 &1 &a\\
J013403.02+304410.6  &23.5125833   &+30.7362778 &0.15 & &15.412 &0.170 &--358 &--96 &52 &1 &a\\
J013406.71+303631.2 &23.5279583  &+30.6086667 &0.16 & &18.145 &-0.018 &--299 &--133 &10 &1 &a\\
J013415.42+302816.4 &23.5642500  &+30.4712222 &0.40 &F0I &17.284 &0.854 &--173 &--29 &20 &2 &a, f\\
J013419.24+303607.3 &23.5801667  &+30.6020278 &0.25 & &18.722 &0.204 &--189 &--12 &13 &1 &a\\
J013419.51+305532.4 &23.5812917  &+30.9256667 &0.48 &A8-F0 &16.905 &0.246 &--254 &9 &16 &1 &a, d\\ 
J013432.80+303942.6 &23.6366667  &+30.6618333 &0.31 &B9I &17.143 &0.044 &--176 &26 &28 &1 &a \\
J013440.89+304619.2 &23.6703750  &+30.7720000 &0.39 &B5I &17.296 &-0.005 &--225 &9 &19 &1 &a\\
J013514.18+304422.6 &23.8090833  &+30.7396111 &0.61 &B5I  &17.654 &-0.038 &--158 &58 &21 &1 &a\\
\hline
J013334.98+303852.4  &23.3957500  &+30.6478889 &0.12 & &18.076 &--0.124 &--196 &--14 &54 &1 &a\\	
IFM-B 600                     &23.3964167  &+30.6001111 &0.14 &LBV &16.429 &0.102 &--184 &--2 &68 &1 &a, h\\
IFM-B 615                     &23.3977083  &+30.5028056 &0.28 &  &18.628 &--0.076 &--110 &72 &41 &3 &a\\
IFM-B 620                     &23.3982500  &+30.4946667 &0.30 &B1I &18.444 &0.042 &--155 &27 &40 &1 &a\\
J013337.09+303521.6  &23.4045417  &+30.5893333 &0.15 &A5I &17.937 &0.062 &--188 &--6 &55 &1 &a, g\\
IFM-B 665                     &23.4083333  &+30.5515556 &0.20 &  &18.538 &--0.048 &--146 &36 &63 &2 &a\\
IFM-B 727                     &23.4153333  &+30.5191389 &0.25 &B5I &17.566	&0.030 &--159 &23 &62 &1 &a, c\\
IFM-B 738                     &23.4163333  &+30.6247778 &0.09 &  &18.863	&--0.085 &--249 &--67 &45 &1 &a\\
J013340.09+302846.1  &23.4170417  &+30.4794722 &0.32 &  &16.047 &0.118 &--206 &--24 &84 &1 &a\\
IFM-B 750                     &23.4181250  &+30.7328889 &0.16 &  &17.697 &--0.091 &--266 &--84 &72 &1 &a\\
J013340.47+303503.3  &23.4186250  &+30.5842500 &0.14 &B9  &17.997 &0.025 &--169 &13 &61 &1 &a, g\\
IFM-B 767                     &23.4189583  &+30.5329722 &0.22 &B6I &18.047 &--0.039 &--156 &26 &44 &1 &a, g\\
IFM-B 770                     &23.4195417  &+30.6832778 &0.09 &B5I &18.264 &--0.037 &--215 &--33 &45 &1 &a\\
J013340.84+303822.5  &23.4201667  &+30.6395833 &0.08 &A2 &18.235 &0.172 &--166 &16 &29 &1 &a, g\\
J013341.36+303629.6  &23.4223333  &+30.6082222 &0.11 &A2 &17.913 &0.106 &--189 &--7 &57 &1 &a, g\\
IFM-B 814                     &23.4269167  &+30.7030833 &0.11 &  &18.735 &0.053 &--245 &--63 &44 &1 &a\\
IFM-B 845                     &23.4302500  &+30.5316111 &0.22 &A0  &17.221 &0.046 &--197 &15 &62 &1 &a, g\\
J013344.27+304247.2  &23.4344583  &+30.7131111 &0.11 &B7I &18.209 &0.011 &--242 &--60 &52 &1 &a, g\\
J013344.43+303843.9  &23.4351250  &+30.6455278 &0.05 &B9  &17.941 &0.054 &--166 &16 &54 &1 &a, g\\
IFM-B 880                     &23.4354167  &+30.6794167 &0.06 & &18.823	 &--0.059 &--111 &71 &35 &3 &a\\
J013344.81+303217.8  &23.4367083  &+30.5382778 &0.21 &B8 &18.067 &0.025 &--175 &7 &54 &1 &a, g\\
J013346.16+303448.5  &23.4423333  &+30.5801389 &0.14 & &17.075 &0.186 &--145 &37 &61 &2 &a\\
IFM-B 963                     &23.4481250  &+30.7297222 &0.13 & &18.641	&--0.088 &--247 &--65 &43 &1 &a\\
IFM-B 1004                   &23.4550833 &+30.6359167 &0.04 &LBV   &16.208	&0.035 &--271 &--89 &69 &1 &a, h\\
IFM-B 1026                   &23.4587083 &+30.6906944 &0.06 &LBV   &16.819	&0.035 &         &    & & &a, i\\
IFM-B 1072                   &23.4633333 &+30.5401389 &0.21 &B9I  &17.194 &0.034 &--178 &4 &68 &1 &a, g\\
IFM-B 1081                   &23.4648333 &+30.6681111 &0.01 &B9.5I  &17.102	&0.039 &--269 &--87 &65 &1 &a\\
IFM-B 1113                   &23.4673333 &+30.5593056 &0.18 &  &17.959 &0.095 &--194 &--12 &60 &1 &a\\
IFM-B 1186                   &23.4800417 &+30.5749722 &0.16 &B5I  &17.572 &--0.009 &--192 &--10 &56 &1 &a\\
IFM-B 1217                   &23.4859583 &+30.5545556 &0.20 &  &17.037 &--0.043 &--204 &--22 &71 &1 &a\\
\enddata
\tablecomments{Parameters of M 31: $R_{25}=95.30$ arcmin \citep{1991rc3..book.....D}, position angle PA = 37.7$^{\rm{o}}$ \citep{1981ApSS..76..477H}, inclination i = 77.5$^{\rm{o}}$ \citep{1978AA....67...73S}. Parameters of M 33: $R_{25}=35.40$ arcmin \citep{1991rc3..book.....D}, position angle PA = 22$^{\rm{o}}$, inclination $i = 53^{\rm{o}}$. 
References: a \citet{2016AJ....152...62M}; b \citet{2019ApJ...884L...7H}; c \citet{2006AJ....131.2478M}; d \citet{2016ApJ...825...50G}; e \citet{1994AA...287..885H}; f \citet{2012ApJ...750...97D}; g \citet{2009ApJ...704.1120U}; h \citet{1953ApJ...118..353H}; i \citet{1996ApJ...469..629M}} 
\end{deluxetable*}

\section{Spectral analysis} \label{sec:analysis}

\subsection{Spectral Analysis Method}
The goal of the spectral anlysis is to obtain effective temperate ($T_{\rm{eff}}$), gravity (log $g$) and metallicity ([$Z$]=log($Z/Z_{\odot}$)). The method is described in detail in \citet{2013ApJ...779L..20K, 2014ApJ...788...56K} and \citet{2014ApJ...785..151H}. We compare normalized observed spectra with synthetic spectra calculated from a comprehensive grid of metal line-blanketed model atmospheres with extensive non-LTE formation calculations \citep{2006A&A...445.1099P, 2008ApJ...681..269K}. The analysis is carried out in several steps. In the first step, we fit the higher Balmer lines  (H$_{4,5,6,7,8,9,10}$) and determine log $g$ at a fixed $T_{\rm{eff}}$ and [$Z$]. This defines a fit curve in the ($T_{\rm{eff}}$, log $g$) diagram along which the calculated Balmer lines are agreement with the observations (see Figure 3 in \citealp{2014ApJ...788...56K}). Then in the second step, 13 spectral windows in the range of 3880 to 6000 \AA~dominated by metal lines are selected to determine $T_{\rm{eff}}$ and [$Z$] by comparing observed and synthetic metal line spectra. The synthetic spectra are used as a function of metallicity for each point along the temperature--gravity relationship to calculate  $\chi^{2}$($T_{\rm{eff}}$, [$Z$]). Using the minimum of $\chi^{2}$($T_{\rm{eff}}$, [$Z$]) and iscontours $\Delta \chi^{2}$($T_{\rm{eff}}$, [$Z$]) around the minimum we determine $T_{\rm{eff}}$ and [$Z$] together with their uncertainties \citep{2014ApJ...788...56K}. In a third step we fit the Balmer lines again as what we did in the first step and determine the gravity, but now with the fixed $T_{\rm{eff}}$ and [$Z$] from the second step. The gravities from the first and third steps are usually slightly different. We then repeat the second step and determine  new $T_{\rm{eff}}$ and [$Z$] by reanalyzing the selected spectral windows again.

\begin{figure*}[ht!]
\centering
\includegraphics[width=0.8\textwidth]{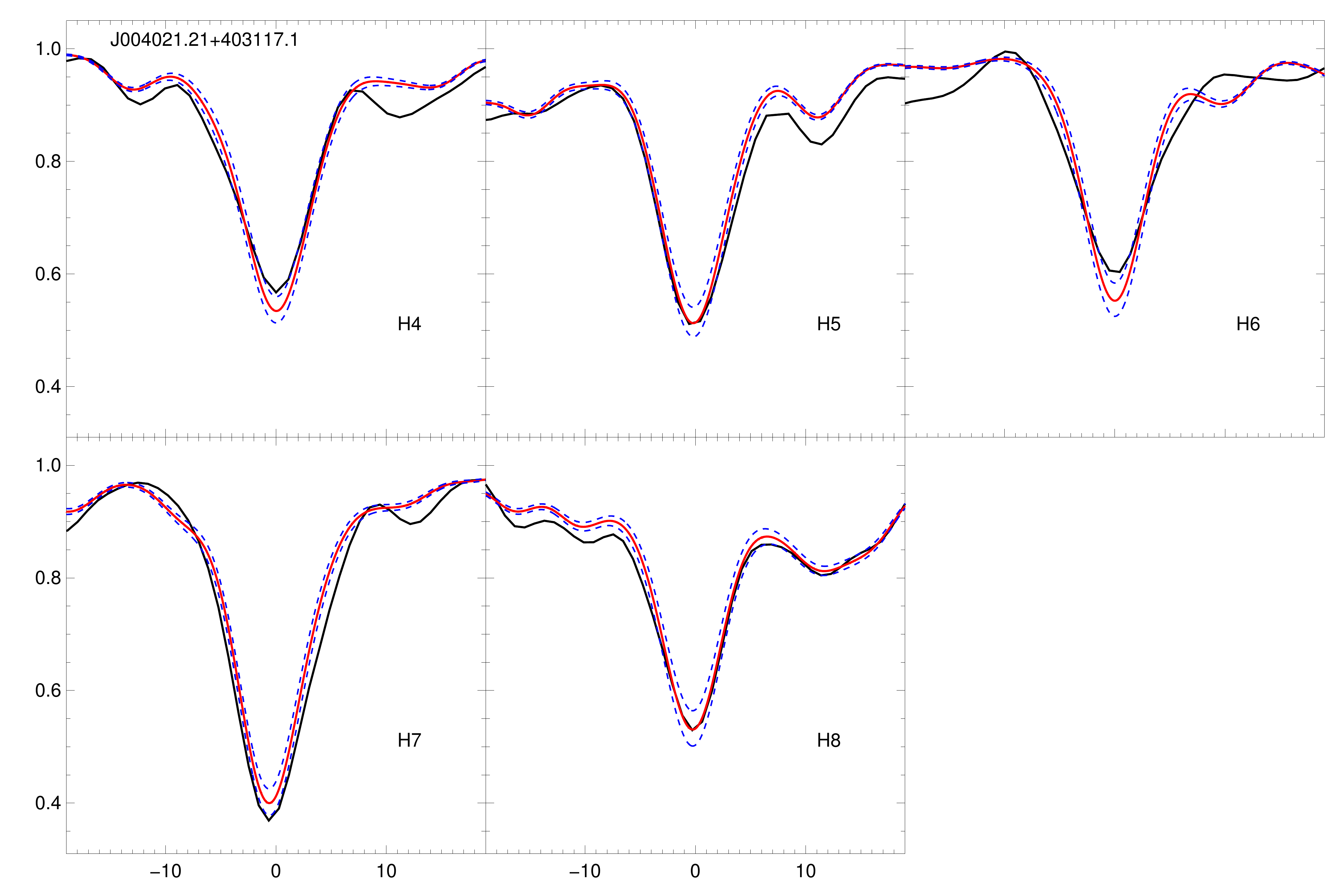}
\includegraphics[width=0.8\textwidth]{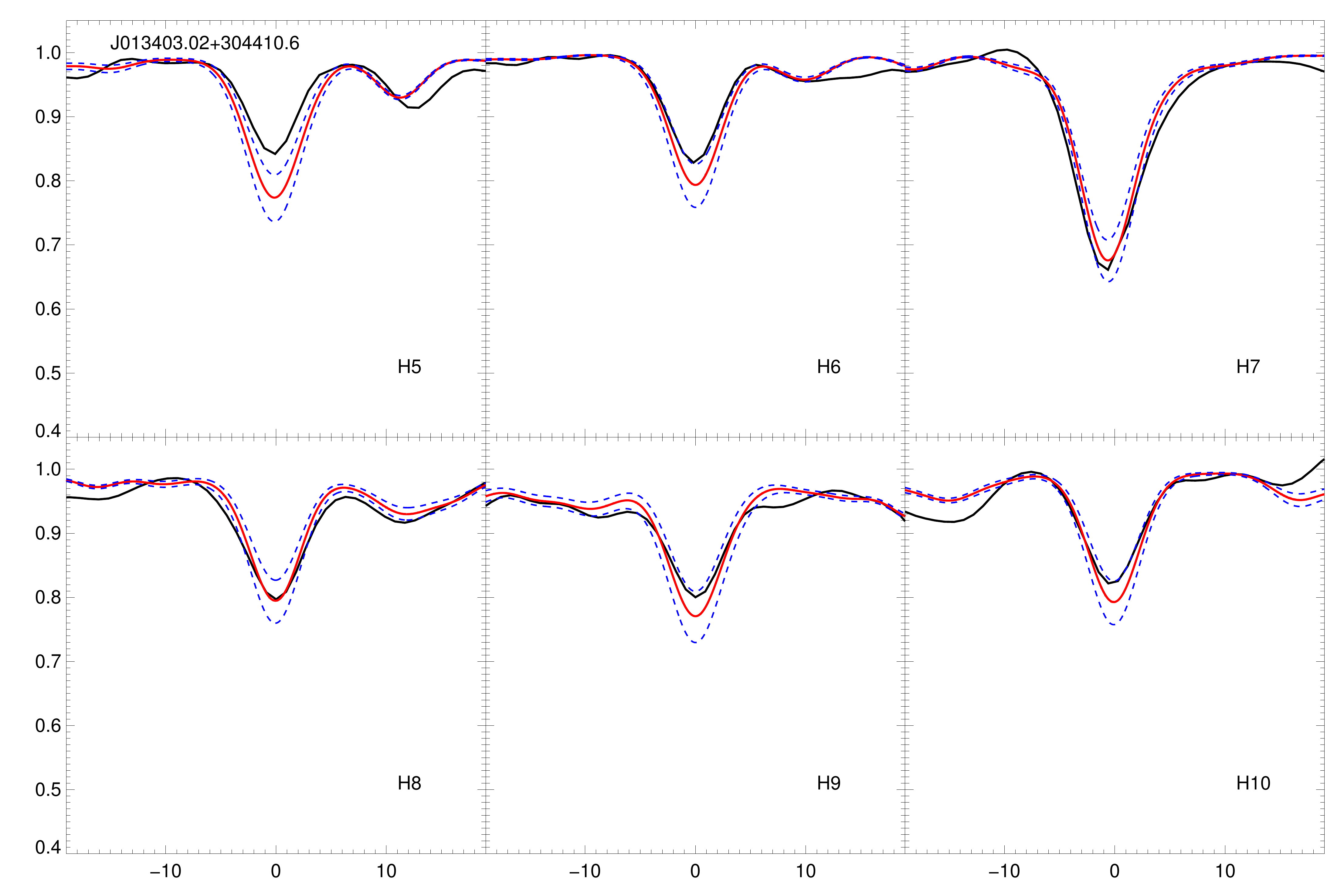}
\caption{Example model fits of the observed Balmer lines profiles for stars from M 31 (top) and M 33 (bottom). The black line is the observed LAMOST spectrum, the red line the best-fit model, and the blue dashed lines are the best-fit models with log g increased and decreased by 0.05 dex. \label{fig:star14_5}}
\end{figure*}

\begin{figure*}[ht!]
\centering
\includegraphics[width=0.8\textwidth]{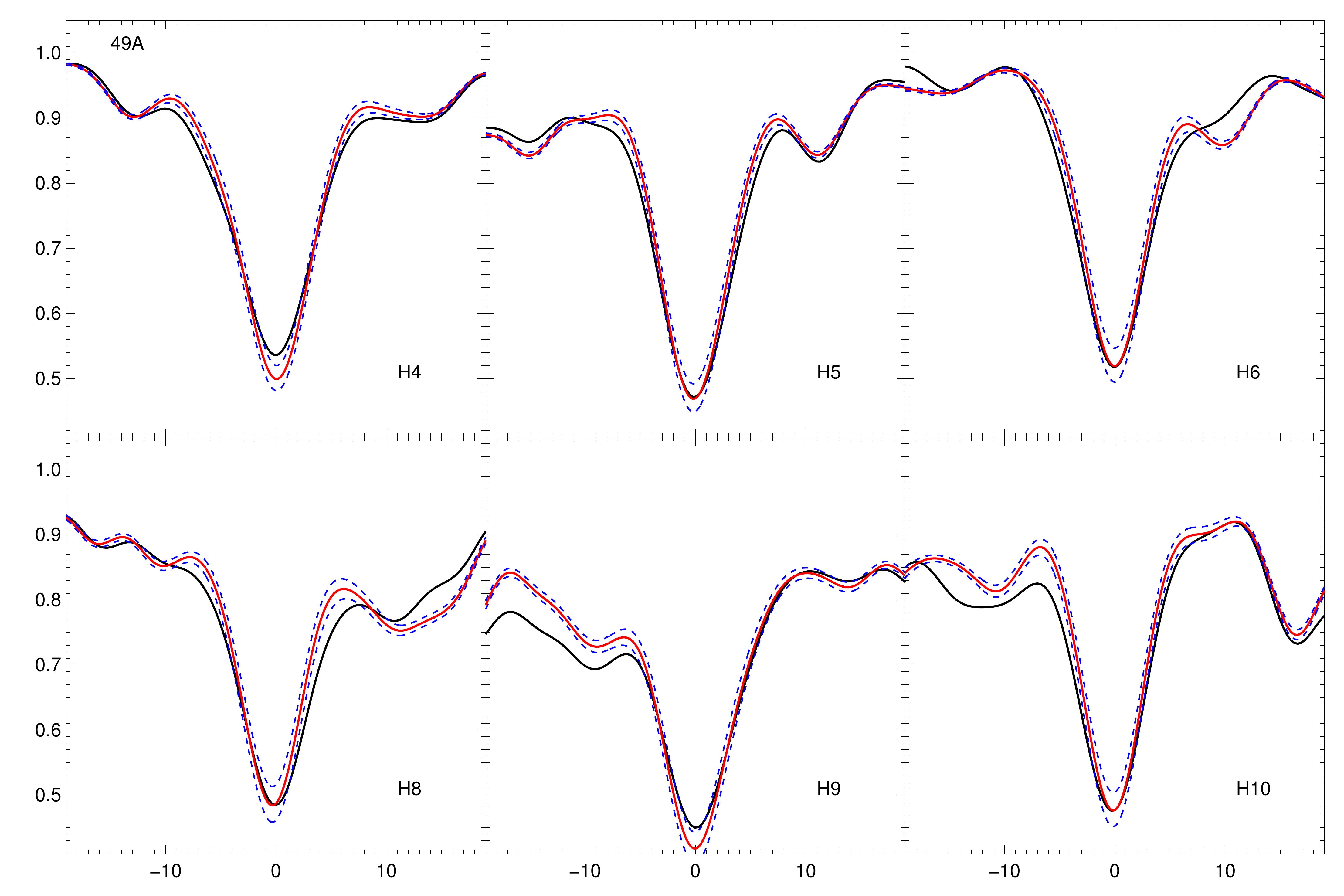}
\caption{Balmer line fits for star 49A, from M 33. The black line is the observed Keck DEIMOS spectrum, the red line the best-fit model, and the blue dashed lines are the best-fit models with log g increased and decreased by 0.05 dex. \label{fig:49A}}
\end{figure*}

To illustrate the fitting of the Balmer lines three examples are given in Figure \ref{fig:star14_5} and Figure \ref{fig:49A}. In Figure \ref{fig:star14_5}, we show two LAMOST targets J004021.21+403117.1 and J013403.02+304410.6 from M 31 and M 33, respectively, while the Keck DEIMOS target 49A (J013346.16+303448.5) from M33 is shown in Figure \ref{fig:49A}.
As in our previous work we note that H$_{4}$ is sometimes contaminated by stellar winds and surrounding H {\scriptsize $\mathrm{II}$} region emission and H$_{7}$ can sometimes be blended with interstellar Ca {\scriptsize $\mathrm{II}$} depending on the strength of interstellar absorption. In extreme cases, we find that the stellar winds and H {\scriptsize $\mathrm{II}$} region emission can also effect H$_{5}$ and H$_{6}$ (e.g. J013403.02+304410.6 in Figure \ref{fig:star14_5}). In those cases, the higher Balmer lines are used to constrain gravity rather than H$_{4}$ and H$_{5}$ which indicate lower gravities. The fit of gravity is typically good to about 0.05--0.1 dex at a fixed temperature (see Figure \ref{fig:star14_5} and \ref{fig:49A}). In some cases we have to abandon the H$_{9}$ and H$_{10}$ lines, because the SNR of LAMOST rapidly decreases towards the shortest wavelengths. In addition, for some of the Keck DEIMOS  H$_{9}$ and H$_{10}$ could not be used, because of the short wavelength range was cut off.

\begin{figure*}[ht!]
\centering
\includegraphics [height=5.3cm,width=4.9cm]{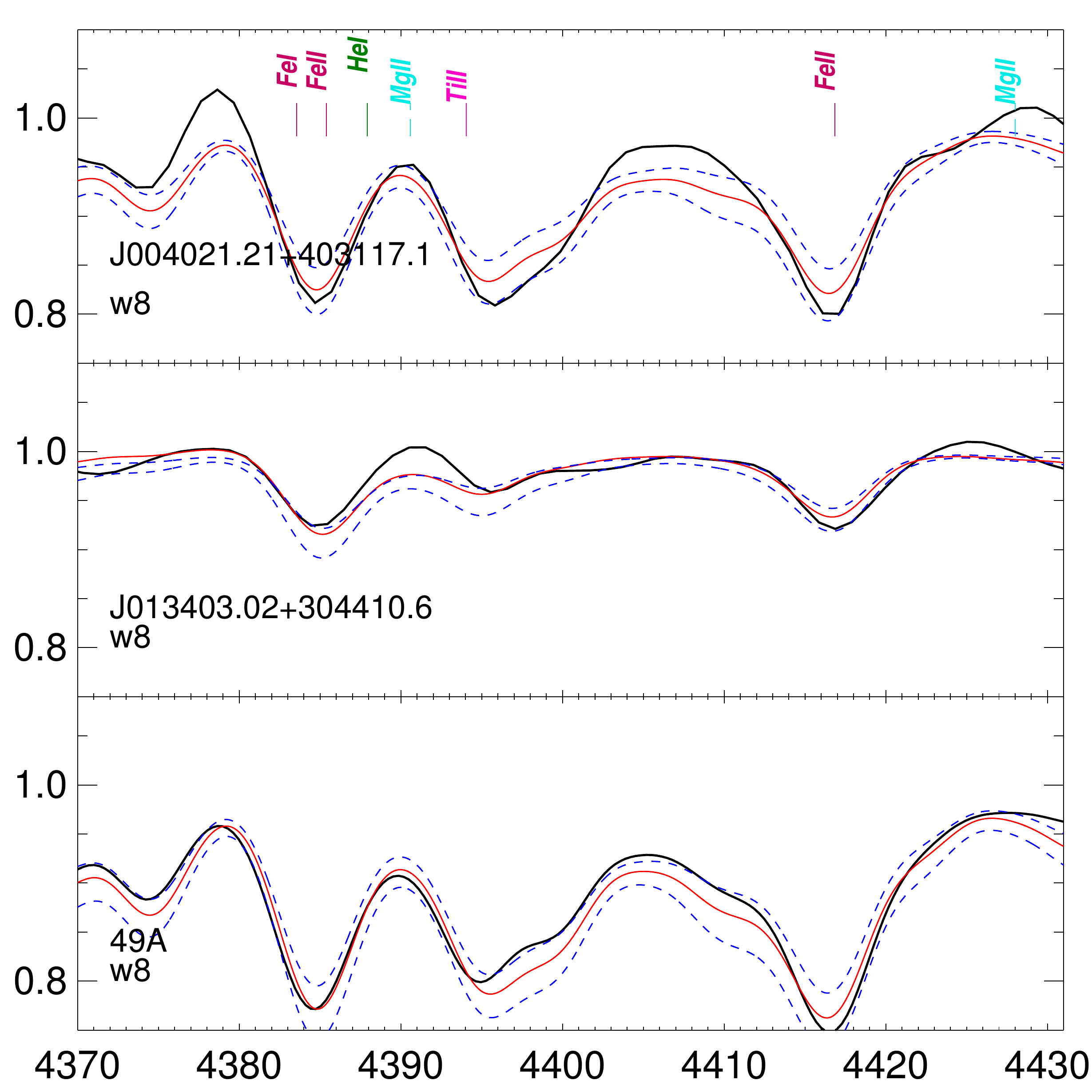}
\includegraphics [height=5.3cm,width=7.9cm]{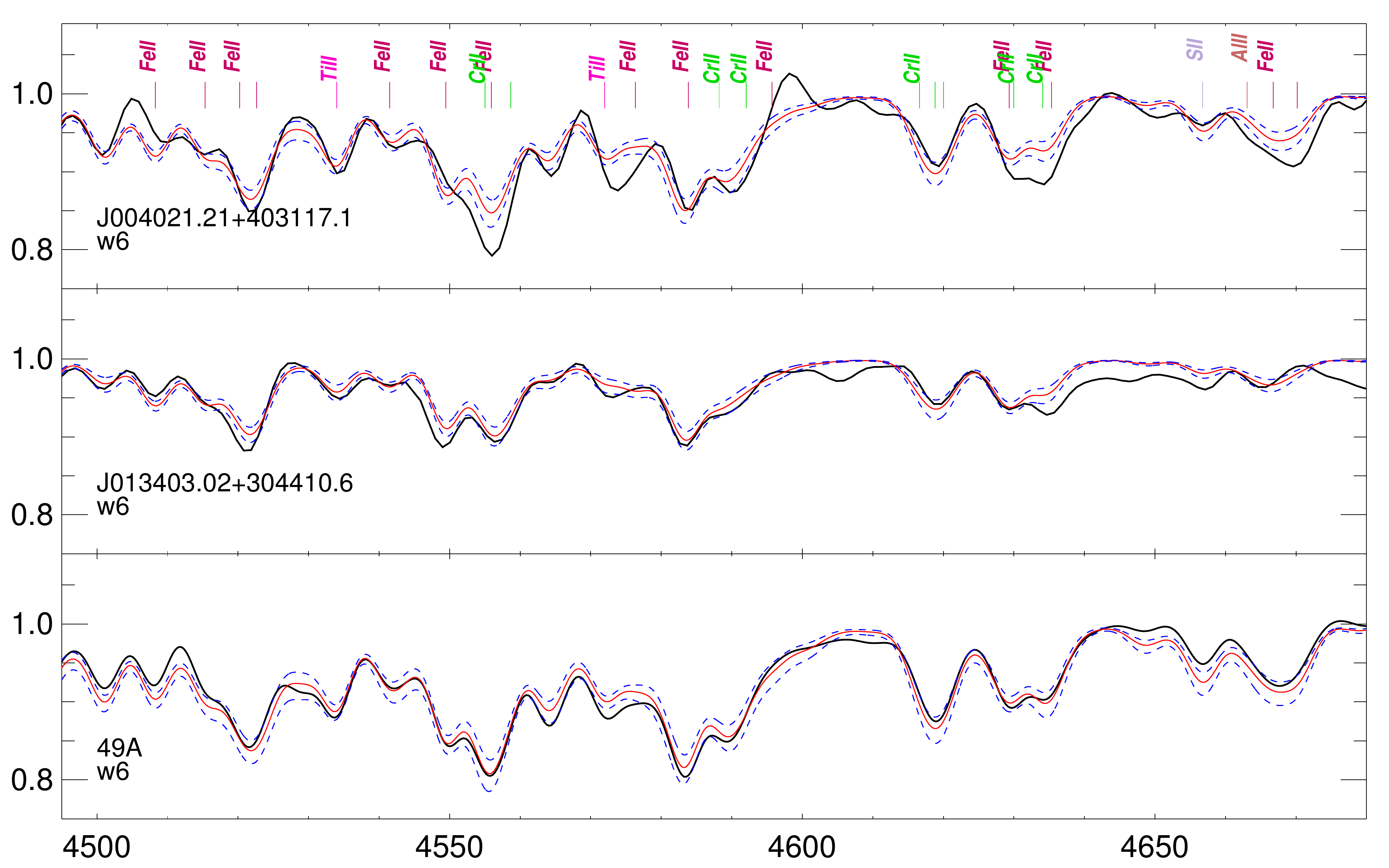}
\includegraphics [height=5.3cm,width=3.9cm]{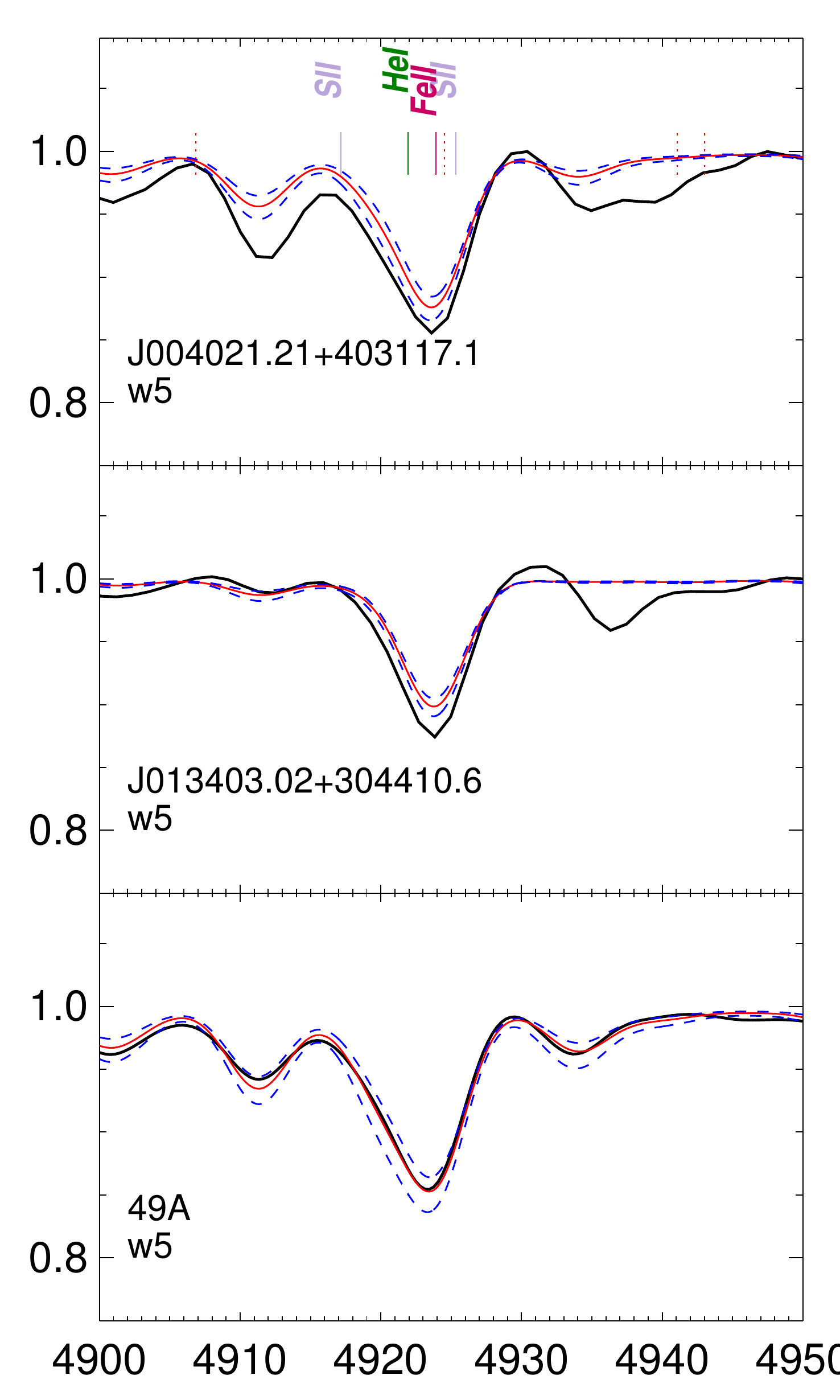}
\includegraphics [height=5.4cm,width=3.7cm]{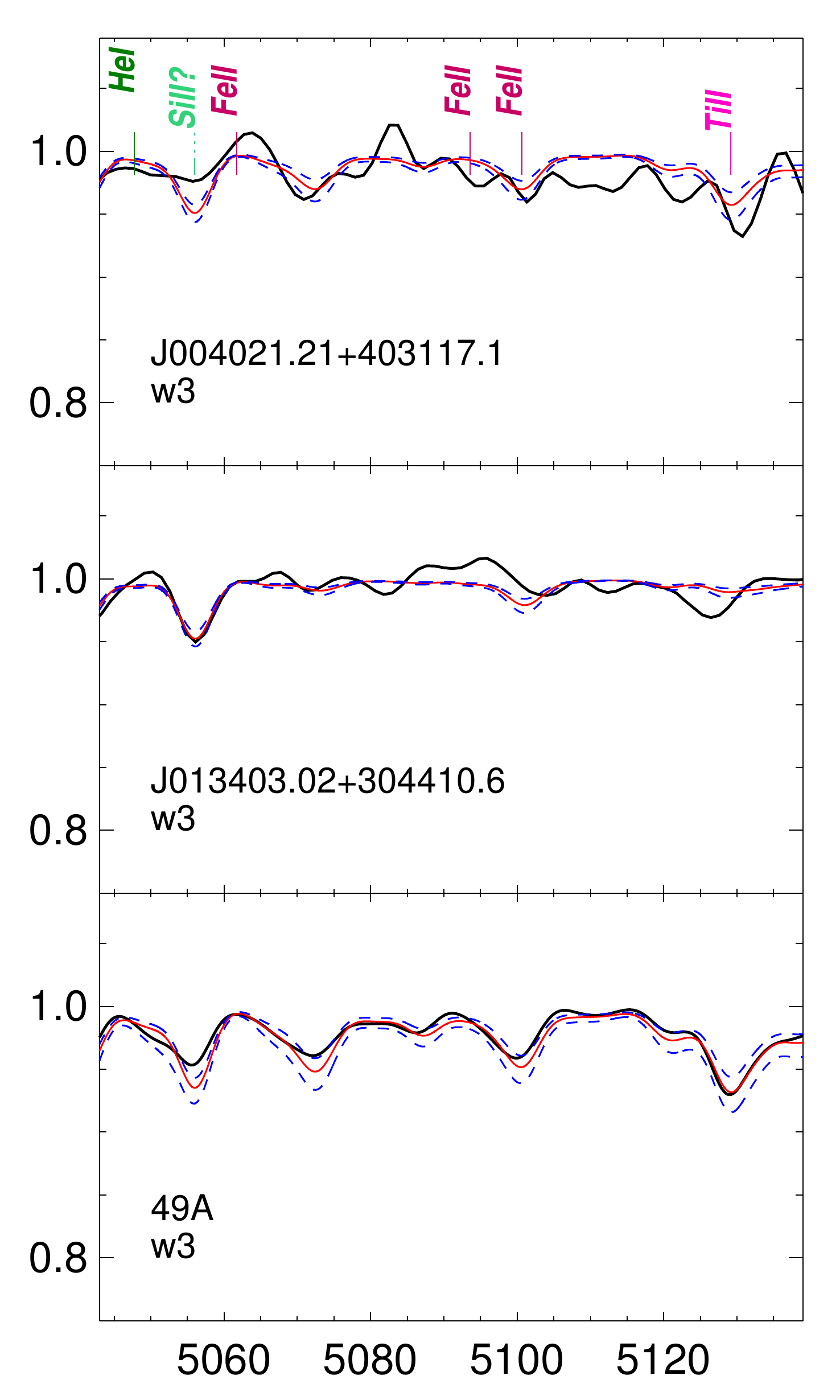}
\includegraphics [height=5.6cm,width=10.5cm]{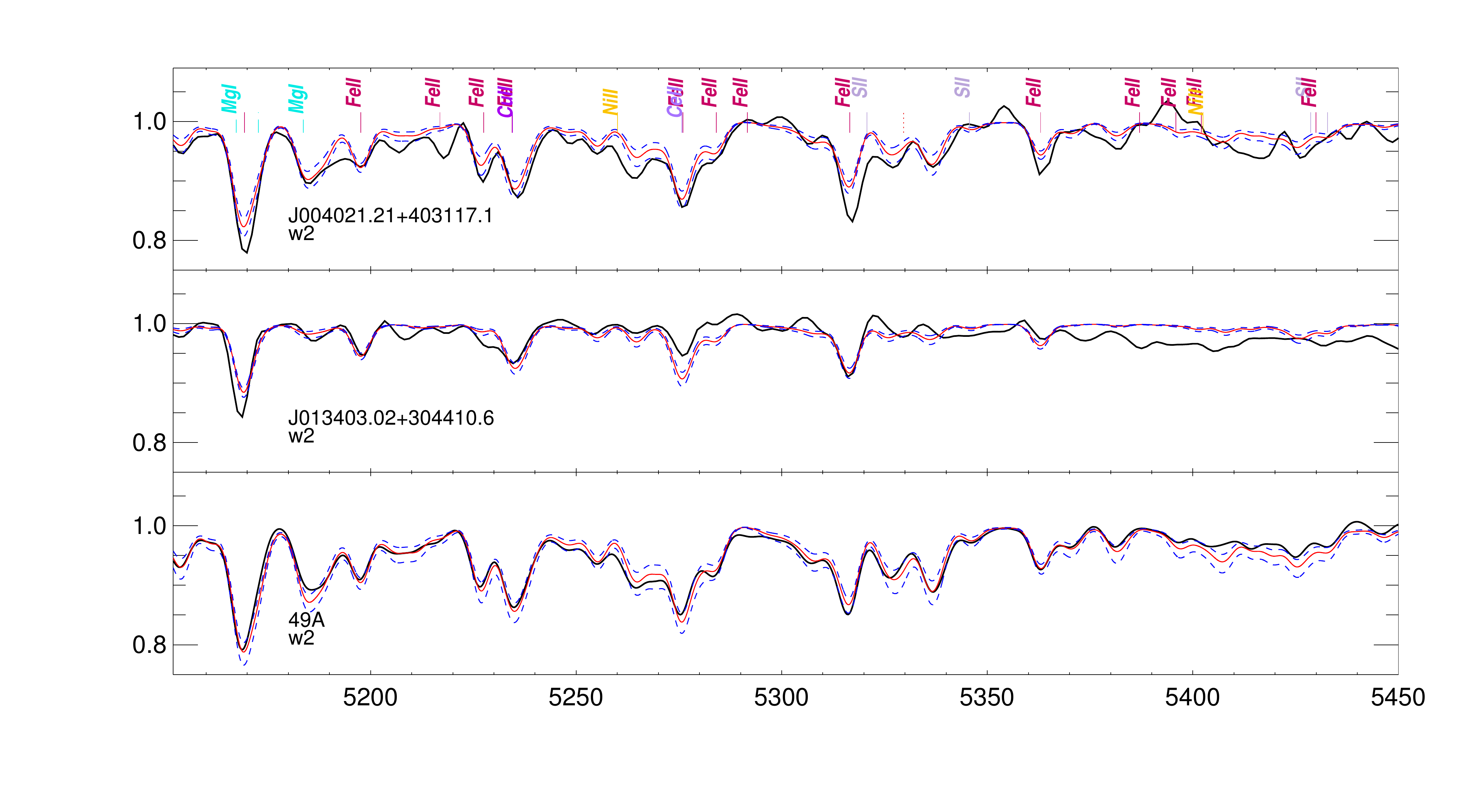}

\caption{Observed metal line spectra of three stars J004021.21+403117.1 (top) from M 31, J013403.02+304410.6 (middle) and 49A (bottom) from M 33 in five spectral windows compared with model calculations (red curves) obtained for the final stellar parameters. The blue dashed lines are the best-fit models with [$Z$] increased and decreased by 0.15 dex. \label{fig:sp_s11_s5_w}}
\end{figure*}

An example of the fit of the metal lines for the targets of Figure \ref{fig:star14_5} and Figure \ref{fig:49A} is given in Figure \ref{fig:sp_s11_s5_w}.

\subsection{Results} \label{Results}
The results of the spectral analysis are given in Table \ref{M31_sp} and \ref{M33_sp}. We find that only 17 out of 43 stars in M 31 sample and only 18 out of 41 stars in M 33 sample have reliable stellar parameters. Firstly, this is because of relative low quality of the LAMOST spectra. Even after the convolution to FWHM = 5 \AA~many target stars have a SNR too low for an accurate quantitative spectral analysis. Secondly, some stars have effective temperatures $T_{\rm{eff}}$ lower than 7900 K, which is outside our model atmosphere grid. These objects are yellow supergiants. Thirdly, although some targets have reasonable spectroscopic parameters, they have abnormally high absolute bolometric magnitudes. For example, with a distance modulus of $\mu=24.51\pm0.13$ mag as determined for M 31 in Section \ref{sect:dist}, the absolute bolometric magnitude of three targets (J004329.02+414127.7, J004413.04+420221.0, and J004535.23+413600.5) are smaller than --10.3 mag. In the M 33 sample  three targets (J013241.52+302408.6, J013251.13+302540.8, and J013307.46+301210.1) have absolute bolometric magnitude $<-10.9$ mag with the distance modulus of $\mu=24.93\pm0.07$ mag obtained in this work (see below). We eliminate these objects for  the measurement of metallicity, metallicity gradient and distance modulus, as  they are very likely contaminated by unresolved neighbor objects.

Some of our objects have been subject to quantitative spectroscopic studies in previous work. J004510.04+413675.6 in M31  has been analyzed by \cite{2000ApJ...541..610V} and \cite{2008mru..conf..332P}, respectively. Our results agree within the error margins. As for M 33, 12 of our targets have been studied by \cite{2009ApJ...704.1120U}. While for ten of these objects the agreement is reasonable, we find somewhat discrepant $T_{\rm{eff}}$ values for two of them, IFM-B 767 and J013344.27+304247.2 ( objects 9 and 6 in Table 2 of U et al.). Our effective temperatures are cooler by about 2000K. We note that U et al. used the Si ionization equilibrium for these two objects to constrain the effective temperature, whereas our method includes lines from additional metals and helium. Because of this discrepancy we regard these objects as uncertain and we do not include them in the further discussion.

\begin{deluxetable*}{crrrrrrrrrrrrr}
\tablecaption{Physical parameters of BSGs in M 31 \label{M31_sp}}
\tablewidth{0pt}
\tablehead{
\colhead{ID} & \colhead{$T_{\rm{eff}}$} & \colhead{log $g$} & \colhead{log $g_{F}$} & \colhead{[Z]} & \colhead{$E(B-V)$} & \colhead{BC} & \colhead{m$_{bol}$}  &\colhead{log$L/L_{\odot}$} &\colhead{R} &\colhead{$M_{\rm{spec}}$} &\colhead{$M_{\rm{evol}}$}\\
\colhead{} & \colhead{K} & \colhead{cgs} & \colhead{cgs} & \colhead{(dex)} & \colhead{(mag)} & \colhead{(mag)} & \colhead{(mag)} &\colhead{(dex)} &\colhead{($R_{\odot}$)} &\colhead{($M_{\odot}$)} &\colhead{($M_{\odot}$)}
}
\startdata
J003720.65+401637.7 & 8250$_{-150}^{+100}$  &0.87$\pm$0.10 &1.20$_{-0.10}^{+0.10}$ &0.00$_{-0.14}^{+0.09}$  &0.14 &0.01 &15.43$\pm$0.11 &5.53$\pm$0.04 &284$\pm$9 &22$\pm$5 &25$\pm$1\\
J003728.99+402007.8 &9750$_{-550}^{+350}$  &1.45$\pm$0.10  &1.49$_{-0.12}^{+0.14}$ &--0.26$_{-0.17}^{+0.26}$ &0.12 &--0.25 &16.69$\pm$0.17 &5.02$\pm$0.07 &114$\pm$11 &13$\pm$4 &16$\pm$1\\
J004005.02+403242.2 &9000$_{-450}^{+400}$  &1.30$\pm$0.05  &1.48$_{-0.13}^{+0.13}$ &0.20$_{-0.15}^{+0.20}$ &0.04 &--0.07 &17.71$\pm$0.16 &4.61$\pm$0.06 &84$\pm$14 &5$\pm$2 &12$\pm$1\\
J004021.21+403117.1  &8050$_{-150}^{+100}$  &0.91$\pm$0.05  &1.29$_{-0.05}^{+0.06}$ &0.00$_{-0.11}^{+0.08}$  &0.19 &0.07 &16.12$\pm$0.06 &5.25$\pm$0.02 &217$\pm$10 &14$\pm$2 &20$\pm$1\\
J004033.90+403047.1 &11150$_{-550}^{+350}$  &1.83$\pm$0.05  &1.64$_{-0.07}^{+0.10}$ &0.03$_{-0.27}^{+0.25}$ &0.18 &--0.48 &16.70$\pm$0.13 &5.02$\pm$0.05 &87$\pm$8 &18$\pm$4 &16$\pm$1\\
J004036.92+410119.0  &9100$_{-400}^{+400}$  &1.29$\pm$0.10  &1.45$_{-0.12}^{+0.13}$ &--0.09$_{-0.26}^{+0.22}$ &0.19 &--0.12 &17.24$\pm$0.15 &4.80$\pm$0.06 &101$\pm$13 &7$\pm$2 &14$\pm$1\\
J004051.59+403303.0 &11150$_{-450}^{+350}$   &1.52$\pm$0.05  &1.33$_{-0.07}^{+0.09}$ &--0.13$_{-0.21}^{+0.17}$ &0.26 &--0.52&15.66$\pm$0.11 &5.43$\pm$0.04 &140$\pm$8 &24$\pm$4 &23$\pm$1\\
J004126.23+405214.3  &9200$_{-500}^{+350}$  &1.25$\pm$0.10  &1.39$_{-0.12}^{+0.14}$ &0.34$_{-0.14}^{+0.13}$ &0.29 &--0.10 &16.48$\pm$0.16 &5.11$\pm$0.06 &141$\pm$8 &13$\pm$4 &17$\pm$1\\
J004133.62+404208.4  &10300$_{-700}^{+400}$  &1.51$\pm$0.10  &1.46$_{-0.08}^{+0.13}$ &0.31$_{-0.48}^{+0.12}$ &0.06 &--0.30 &17.35$\pm$0.18 &4.76$\pm$0.07 &75$\pm$11 &6$\pm$2 &13$\pm$1\\
J004154.82+405706.7 &10800$_{-350}^{+350}$  &1.76$\pm$0.05  &1.63$_{-0.08}^{+0.08}$ &0.07$_{-0.18}^{+0.22}$ &0.23 &--0.41 &16.80$\pm$0.10 &4.98$\pm$0.04 &89$\pm$8 &16$\pm$3 &16$\pm$1\\
J004202.86+411434.6  &9150$_{-350}^{+500}$   &1.36$\pm$0.05  &1.51$_{-0.11}^{+0.08}$ &0.21$_{-0.08}^{+0.19}$  &0.25 &--0.10 &17.24$\pm$0.12 &4.81$\pm$0.05 &101$\pm$13 &9$\pm$2 &14$\pm$1\\
J004212.27+413527.4  &8500$_{-500}^{+400}$   &1.20$\pm$0.10  &1.48$_{-0.16}^{+0.12}$ &0.20$_{-0.24}^{+0.27}$ &0.18 &0.02 &17.00$\pm$0.18 &4.90$\pm$0.06 &130$\pm$16 &10$\pm$4 &15$\pm$1\\
J004253.42+412700.5  &10650$_{-650}^{+550}$   &1.73$\pm$0.10  &1.62$_{-0.10}^{+0.11}$ &0.14$_{-0.24}^{+0.21}$ &0.06 &--0.38 &16.68$\pm$0.16 &5.03$\pm$0.06 &96$\pm$11 &18$\pm$4 &16$\pm$1\\
J004422.84+420433.1  &9850$_{-650}^{+650}$  &1.48$\pm$0.10  &1.51$_{-0.15}^{+0.15}$ &--0.10$_{-0.25}^{+0.25}$ &0.03 &--0.26 &16.10$\pm$0.20 &5.26$\pm$0.08 &147$\pm$13 &24$\pm$7 &20$\pm$1\\
J004510.04+413657.6 &8200$_{-100}^{+150}$  &0.89$\pm$0.05 &1.23$_{-0.06}^{+0.05}$ &0.10$_{-0.11}^{+0.10}$ &0.17 &0.03 &15.85$\pm$0.06 &5.36$\pm$0.02 &237$\pm$9 &16$\pm$2 &21$\pm$1\\
J004535.26+413238.6  &8500$_{-450}^{+750}$  &1.00$\pm$0.05  &1.28$_{-0.16}^{+0.10}$ &--0.26$_{-0.25}^{+0.38}$ & 0.54 &--0.04 &15.08$\pm$0.17 &5.67$\pm$0.07 &315$\pm$19 &36$\pm$6 &28$\pm$2\\
J004623.14+413847.5 &10050$_{-450}^{+550}$  &1.36$\pm$0.10  &1.35$_{-0.11}^{+0.11}$ &0.02$_{-0.20}^{+0.24}$ &0.15 &--0.30 &15.38$\pm$0.15 &5.55$\pm$0.06 &196$\pm$12 &32$\pm$5 &25$\pm$2\\
\enddata
\end{deluxetable*}

\begin{deluxetable*}{rrrrrrrrrrrrrr}
\tablecaption{Stellar Parameters of Supergiants in M 33 \label{M33_sp}}
\tablecolumns{10}
\tablewidth{0pt}
\tablehead{
\colhead{ID} & \colhead{$T_{eff}$} & \colhead{log $g$} & \colhead{log $g_{F}$} & \colhead{[Z]} & \colhead{$E(B-V)$} & \colhead{BC} & \colhead{m$_{bol}$} &\colhead{log$L/L_{\odot}$} &\colhead{R} &\colhead{$M_{\rm{spec}}$} &\colhead{$M_{\rm{evol}}$} \\
\colhead{} & \colhead{K} & \colhead{cgs} & \colhead{cgs} & \colhead{(dex)} & \colhead{(mag)} & \colhead{(mag)} & \colhead{(mag)} &\colhead{(dex)} &\colhead{($R_{\odot}$)} &\colhead{($M_{\odot}$)} &\colhead{($M_{\odot}$)}
}
\startdata
J013242.51+302455.3  &8150$_{-75}^{+175}$   &0.84$\pm$0.10  &1.15$_{-0.10}^{+0.10}$ &0.15$_{-0.04}^{+0.07}$ &0.14 &0.01 &14.97$\pm$0.10 &5.88$\pm$0.04 &437$\pm$8 &43$\pm$10 &35$\pm$2\\
J013300.23+302323.7  &8500$_{-150}^{+100}$   &0.95$\pm$0.10 &1.23$_{-0.10}^{+0.10}$ &-0.12$_{-0.09}^{+0.09}$ &0.07 &-0.04 &16.18$\pm$0.11 &5.40$\pm$0.04 &230$\pm$9 &17$\pm$4 &22$\pm$1\\
J013322.43+303513.2 &11400$_{-400}^{+400}$   &1.93$\pm$0.10 &1.70$_{-0.12}^{+0.12}$ &-0.14$_{-0.19}^{+0.06}$ &0.01 &-0.54 &17.79$\pm$0.10 &4.75$\pm$0.04 &61$\pm$7 &12$\pm$3 &13$\pm$1\\
J013323.57+302221.8  &8300$_{-200}^{+300}$   &0.90$\pm$0.05 &1.22$_{-0.08}^{+0.07}$ &-0.36$_{-0.19}^{+0.16}$ &0.52 &-0.03 &15.00$\pm$0.09 &5.87$\pm$0.04 &417$\pm$13 &50$\pm$7 &35$\pm$2\\
J013324.14+300520.9  &8500$_{-200}^{+250}$   &0.95$\pm$0.10 &1.23$_{-0.11}^{+0.11}$ &-0.14$_{-0.23}^{+0.31}$ &0.54 &-0.04 &15.14$\pm$0.09 &5.81$\pm$0.04 &373$\pm$11 &45$\pm$6 &33$\pm$2\\
J013337.09+303521.6  &8600$_{-300}^{+300}$   &1.26$\pm$0.10  &1.52$_{-0.12}^{+0.12}$ &0.12$_{-0.18}^{+0.17}$ &0.03 &0.00 &17.83$\pm$0.14 &4.74$\pm$0.06 &105$\pm$13 &7$\pm$1 &13$\pm$1\\
J013344.66+303631.6  &9150$_{-500}^{+550}$   &1.36$\pm$0.10  &1.51$_{-0.14}^{+0.14}$ &0.37$_{-0.17}^{+0.13}$ &0.02 &-0.08 &16.75$\pm$0.18 &5.17$\pm$0.07 &153$\pm$15 &19$\pm$2 &18$\pm$1\\
J013351.20+303224.5 &10500$_{-400}^{+300}$   &1.55$\pm$0.10 &1.47$_{-0.11}^{+0.12}$ &-0.02$_{-0.23}^{+0.27}$ &0.07 &-0.37 &16.60$\pm$0.14 &5.22$\pm$0.05 &124$\pm$8 &20$\pm$5 &19$\pm$1\\
J013356.89+300900.6  &8300$_{-200}^{+300}$   &1.05$\pm$0.10 &1.22$_{-0.12}^{+0.10}$ &-0.40$_{-0.10}^{+0.15}$ &0.47 &-0.03 &16.31$\pm$0.09 &5.35$\pm$0.04 &228$\pm$13 &15$\pm$2 &35$\pm$2\\
J013359.74+304124.4 &10000$_{-250}^{+300}$   &1.30$\pm$0.10  &1.30$_{-0.11}^{+0.11}$ &0.34$_{-0.07}^{+0.10}$ &0.04 &-0.26 &17.02$\pm$0.12 &5.06$\pm$0.05 &113$\pm$8 &9$\pm$3 &17$\pm$1\\
J013403.02+304410.6  &8700$_{-200}^{+100}$   &0.95$\pm$0.05 &1.19$_{-0.05}^{+0.06}$ &0.06$_{-0.07}^{+0.07}$ &0.10 &-0.08 &15.01$\pm$0.07 &5.86$\pm$0.03 &376$\pm$9 &46$\pm$6 &34$\pm$2\\
J013406.71+303631.2  &11600$_{-800}^{+600}$  &1.82$\pm$0.10  &1.65$_{-0.13}^{+0.16}$ &0.30$_{-0.20}^{+0.20}$ &0.06 &--0.54 &17.41$\pm$0.19 &4.91$\pm$0.08 &70$\pm$9 &12$\pm$4 &15$\pm$1\\
J013415.42+302816.4  &8300$_{-150}^{+200}$   &0.90$\pm$0.10  &1.22$_{-0.11}^{+0.10}$ &0.21$_{-0.10}^{+0.11}$ &0.77 &0.04 &14.93$\pm$0.07 &5.89$\pm$0.03 &429$\pm$11 &60$\pm$7 &35$\pm$2\\
J013419.24+303607.3  &8600$_{-300}^{+300}$   &1.37$\pm$0.10  &1.63$_{-0.12}^{+0.12}$ &0.15$_{-0.15}^{+0.18}$ &0.19 &0.01 &18.16$\pm$0.10 &4.61$\pm$0.04 &91$\pm$13 &7$\pm$2 &12$\pm$1\\
J013419.51+305532.4  &8050$_{-150}^{+150}$   &0.81$\pm$0.10  &1.19$_{-0.11}^{+0.11}$ &0.08$_{-0.12}^{+0.12}$ &0.14 &0.04 &16.52$\pm$0.07 &5.26$\pm$0.03 &219$\pm$11 &11$\pm$2 &20$\pm$1\\
J013432.80+303942.6 &10000$_{-200}^{+400}$   &1.35$\pm$0.05  &1.35$_{-0.09}^{+0.06}$ &0.15$_{-0.08}^{+0.10}$ &0.06 &-0.28 &16.69$\pm$0.08 &5.19$\pm$0.03 &131$\pm$9 &14$\pm$2 &19$\pm$1\\
J013440.89+304619.2 &12500$_{-1000}^{+1000}$   &1.80$\pm$0.05  &1.41$_{-0.15}^{+0.15}$ &0.00$_{-0.17}^{+0.30}$ &0.08 &-0.74 &16.31$\pm$0.21 &5.35$\pm$0.08 &101$\pm$9 &23$\pm$5 &21$\pm$2\\
J013514.18+304422.6 &14000$_{-1000}^{+500}$   &2.00$\pm$0.10  &1.42$_{-0.12}^{+0.16}$ &0.21$_{-0.07}^{+0.10}$ &0.08 &-1.02 &16.40$\pm$0.19 &5.31$\pm$0.08 &77$\pm$6 &21$\pm$6 &20$\pm$1\\
\hline
IFM-B 600  &8550$_{-250}^{+350}$   &1.08$\pm$0.10  &1.35$_{-0.09}^{+0.07}$ &--0.25$_{-0.15}^{+0.17}$ &0.05 &--0.04 &16.23$\pm$0.10 &5.38$\pm$0.04 &223$\pm$13 &22$\pm$4 &22$\pm$1\\
IFM-B 615 &14450$_{-250}^{+300}$   &2.39$\pm$0.10  &1.75$_{-0.11}^{+0.10}$ &--0.02$_{-0.05}^{+0.08}$ &0.06 &--1.07 &17.39$\pm$0.11 &4.91$\pm$0.04 &46$\pm$3 &19$\pm$5 &15$\pm$1\\
J013337.09+303521.6  &8700$_{-150}^{+150}$   &1.37$\pm$0.05   &1.61$_{-0.05}^{+0.06}$ &0.02$_{-0.10}^{+0.10}$ &0.05 &--0.02 &17.76$\pm$0.07 &4.76$\pm$0.03 &106$\pm$9 &10$\pm$2 &14$\pm$1\\
IFM-B 665  &9750$_{-250}^{+450}$   &1.90$\pm$0.05  &1.94$_{-0.09}^{+0.07}$ &--0.23$_{-0.07}^{+0.11}$ &0.00 &--0.24 &18.31$\pm$0.14 &4.54$\pm$0.05 &66$\pm$10 &12$\pm$4 &12$\pm$1\\
  IFM-B 727 &12300$_{-200}^{+400}$  &1.97$\pm$0.10   &1.61$_{-0.11}^{+0.10}$ &--0.03$_{-0.12}^{+0.12}$ &0.12 &--0.69 &16.49$\pm$0.15 &5.27$\pm$0.06 &95$\pm$5 &31$\pm$8 &20$\pm$1\\
J013340.47+303503.3 &11650$_{-250}^{+300}$  &1.78$\pm$0.10  &1.51$_{-0.11}^{+0.11}$ &--0.03$_{-0.09}^{+0.05}$ &0.10 &--0.58 &17.11$\pm$0.12 &5.74$\pm$0.05 &80$\pm$6 &14$\pm$4 &16$\pm$1\\
IFM-B 767 &13700$_{-300}^{+300}$   &2.08$\pm$0.10  &1.53$_{-0.11}^{+0.11}$ &--0.10$_{-0.06}^{+0.06}$ &0.08 &--0.95 &16.85$\pm$0.12 &5.13$\pm$0.05 &65$\pm$4 &19$\pm$5 &18$\pm$1\\
IFM-B 770 &13450$_{-400}^{+300}$  &1.99$\pm$0.10  &1.48$_{-0.11}^{+0.11}$ &--0.03$_{-0.06}^{+0.07}$ &0.07 &--0.90 &17.13$\pm$0.12 &5.02$\pm$0.05 &59$\pm$5 &13$\pm$3 &16$\pm$1\\
J013340.84+303822.5  &8450$_{-150}^{+100}$   &1.41$\pm$0.10   &1.70$_{-0.10}^{+0.10}$ &0.36$_{-0.08}^{+0.07}$ &0.15 &0.06 &17.83$\pm$0.11 &4.74$\pm$0.04 &109$\pm$9 &11$\pm$3 &13$\pm$1\\
J013341.36+303629.6  &8500$_{-50}^{+25}$   &1.40$\pm$0.10  &1.68$_{-0.10}^{+0.10}$ &0.07$_{-0.07}^{+0.07}$ &0.09 &0.02 &17.65$\pm$0.10 &4.81$\pm$0.04 &117$\pm$5 &13$\pm$3 &14$\pm$1\\
IFM-B 814 &12000$_{-100}^{+100}$   &1.95$\pm$0.10   &1.63$_{-0.10}^{+0.10}$ &0.15$_{-0.05}^{+0.04}$ &0.14 &--0.62 &17.67$\pm$0.10 &4.80$\pm$0.04 &58$\pm$3 &11$\pm$3 &14$\pm$1\\
IFM-B 845 &10000$_{-100}^{+100}$   &1.45$\pm$0.10  &1.45$_{-0.10}^{+0.10}$ &--0.13$_{-0.04}^{+0.03}$ &0.07 &--0.29 &16.73$\pm$0.10 &4.18$\pm$0.04 &129$\pm$5 &17$\pm$4 &18$\pm$1\\
J013344.27+304247.2 &13650$_{-350}^{+350}$   &2.07$\pm$0.10   &1.53$_{-0.11}^{+0.11}$ &0.08$_{-0.08}^{+0.08}$ &0.13 &--0.93 &16.88$\pm$0.12 &5.12$\pm$0.05 &65$\pm$4 &18$\pm$5 &18$\pm$1\\
J013344.43+303843.9 &10500$_{-100}^{+200}$   &1.60$\pm$0.05   &1.52$_{-0.06}^{+0.05}$ &0.02$_{-0.04}^{+0.06}$ &0.10 &--0.37 &17.28$\pm$0.06 &5.50$\pm$0.02 &91$\pm$6 &12$\pm$2 &16$\pm$1\\
J013344.81+303217.8 &12500$_{-300}^{+300}$   &1.85$\pm$0.10   &1.46$_{-0.11}^{+0.11}$ &0.07$_{-0.08}^{+0.08}$ &0.12 &--0.73 &16.98$\pm$0.12 &5.08$\pm$0.05 &74$\pm$5 &14$\pm$4 &17$\pm$1\\
 J013346.16+303448.5  &8050$_{-50}^{+50}$   &0.96$\pm$0.05   &1.34$_{-0.05}^{+0.05}$ &0.29$_{-0.03}^{+0.04}$ &0.10 &0.10 &16.85$\pm$0.05 &5.13$\pm$0.02 &188$\pm$6 &12$\pm$2 &18$\pm$1\\
IFM-B 963 &14350$_{-450}^{+350}$  &2.03$\pm$0.10 &1.40$_{-0.11}^{+0.11}$ &--0.15$_{-0.02}^{+0.03}$ &0.03 &--1.08 &17.47$\pm$0.13 &4.88$\pm$0.05 &45$\pm$4 &8$\pm$2 &15$\pm$1\\
IFM-B 1072 &11000$_{-200}^{+100}$   &1.60$\pm$0.10  &1.43$_{-0.10}^{+0.10}$ &--0.15$_{-0.04}^{+0.04}$ &0.08 &--0.48 &16.46$\pm$0.10 &5.28$\pm$0.04 &121$\pm$4 &21$\pm$5 &20$\pm$1\\
IFM-B 1081 &11000$_{-100}^{+150}$   &1.45$\pm$0.10   &1.28$_{-0.10}^{+0.10}$ &0.20$_{-0.06}^{+0.05}$ &0.08 &--0.50 &16.37$\pm$0.10 &5.32$\pm$0.04 &126$\pm$4 &16$\pm$4 &21$\pm$1\\
IFM-B 1113 &14550$_{-300}^{+300}$   &2.00$\pm$0.05   &1.35$_{-0.06}^{+0.06}$ &0.01$_{-0.04}^{+0.04}$ &0.22 &--1.10 &16.18$\pm$0.07 &5.39$\pm$0.03 &79$\pm$3 &22$\pm$3 &22$\pm$1\\
IFM-B 1186 &13600$_{-200}^{+300}$   &1.91$\pm$0.05   &1.38$_{-0.06}^{+0.06}$ &0.29$_{-0.03}^{+0.03}$ &0.10 &--0.91 &16.34$\pm$0.07 &5.33$\pm$0.03 &84$\pm$4 &21$\pm$3 &21$\pm$1\\
IFM-B 1217 &14400$_{-700}^{+500}$  &1.80$\pm$0.05  &1.17$_{-0.08}^{+0.10}$ &--0.20$_{-0.09}^{+0.06}$ &0.05 &--1.14 &15.73$\pm$0.13 &5.58$\pm$0.05 &99$\pm$5 &22$\pm$3 &26$\pm$2\\
\enddata
\end{deluxetable*}

\section{Reddening and stellar properties}\label{sec:red-sp}
\subsection{Interstellar Reddening in M 31}
With the stellar parameters listed in Table \ref{M31_sp} and \ref{M33_sp}, we obtain the intrinsic colors of our stars from the model atmosphere flux distributions \citep[see][]{2006A&A...445.1099P, 2008ApJ...681..269K}. Comparing with the observed colors from LGGS (see Table \ref{M31_M33_sample}) and applying the interstellar extinction law  by \cite{1989ApJ...345..245C}, we calculate interstellar reddening $E(B-V)$. Assuming $A_{V}=3.1E(B-V)$ for the relation between visual extinction and reddening, we then determine dereddened apparent bolometric magnitudes $m_{\rm{bol}}$ by adding the bolometric correction (BC) obtained from the model atmosphere calculation \citep[see][for details]{2008ApJ...681..269K}. Figure \ref{sect5:EBV} shows the distribution of interstellar reddening $E(B-V)$ among the stars in M 31 and M 33 samples. For M 31 sample, the average reddening is $<E(B-V)>=0.18$ mag with a dispersion of 0.11 mag. This is larger than the foreground reddening of 0.062 mag measured by \cite{1998ApJ...500..525S}, but is consistent with the value of 0.20 mag found by \cite{2001ApJ...553...47F} in the $\it{HST}$ key project study of Cepheids. Our result is in close agreement to the typical reddening $E(B-V)=0.13$ mag found by \cite{2007AJ....134.2474M} for massive stars.

Information about interstellar reddening can also be obtained from the Balmer decrement of H {\scriptsize $\mathrm{II}$} regions. \cite{2012MNRAS.427.1463Z} have published reddening values c(H$_{\beta}$) for a group of M 31 H {\scriptsize $\mathrm{II}$} regions. Adopting c(H$_{\beta}$)/$E(B-V)=1.43$ \citep{1985ApJ...297..724K}, we calculate $E(B-V)$ for each H {\scriptsize $\mathrm{II}$} region, and find  a large scatter in reddening from 0.0 to 0.7 mag with an average of $<E(B-V)>=0.36$ mag.  Also using H {\scriptsize $\mathrm{II}$} region Balmer decrements, \cite{2012ApJ...758..133S} detected an even large range of extinction between 0.0 to 1.5 mag depending on the location within the galaxy. They also got a similar result from the study of PNe. Comparing with H {\scriptsize $\mathrm{II}$} regions, it seems like that the massive stars give lower reddening. This is very likely a selection effect, because our massive star targets were selected with respect to brightness and relatively blue colors. 

\subsection{Interstellar Reddening in M 33}
\begin{figure}[ht!]
\centering
\includegraphics[width=0.50\textwidth]{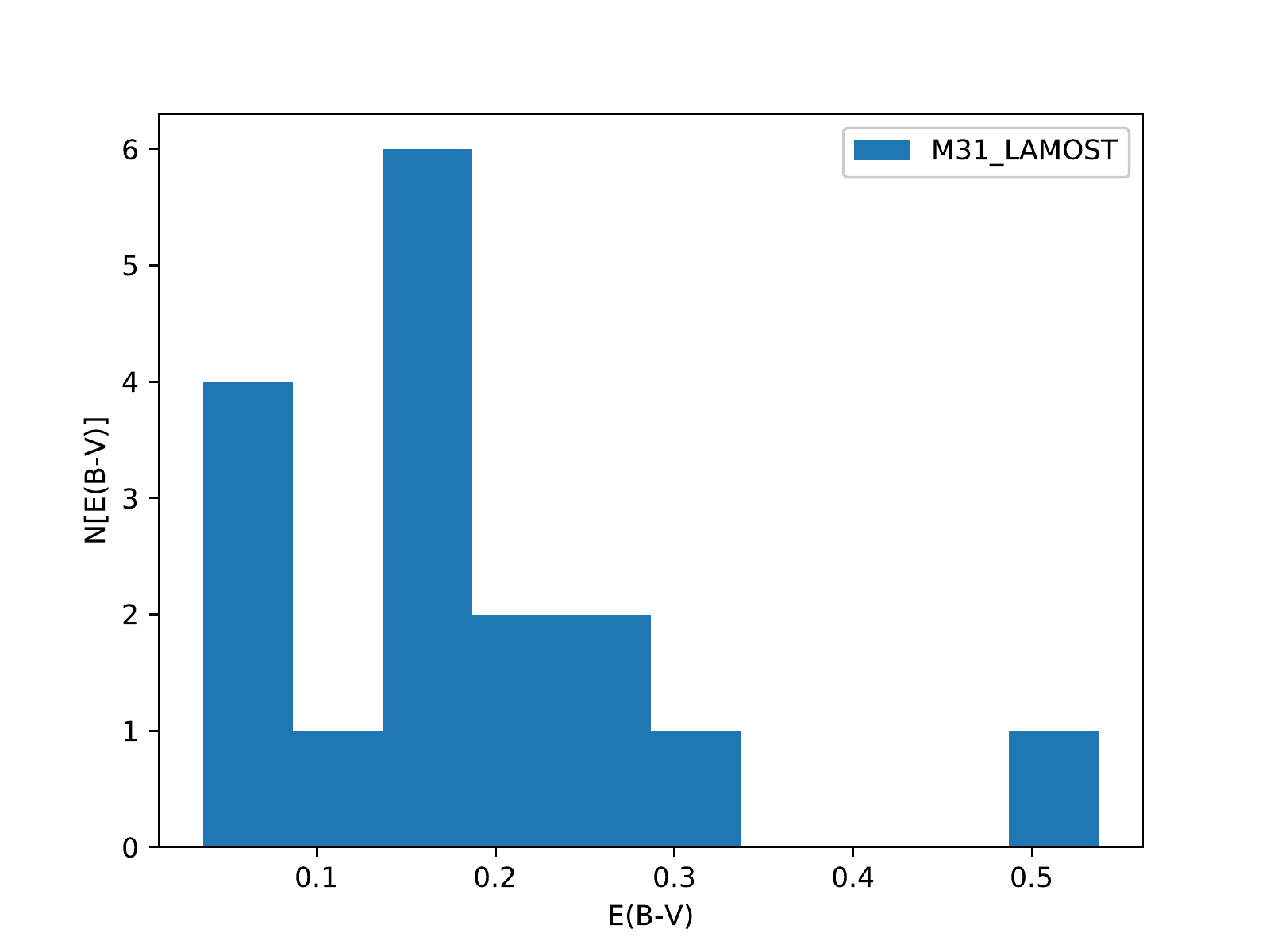}
\includegraphics[width=0.50\textwidth]{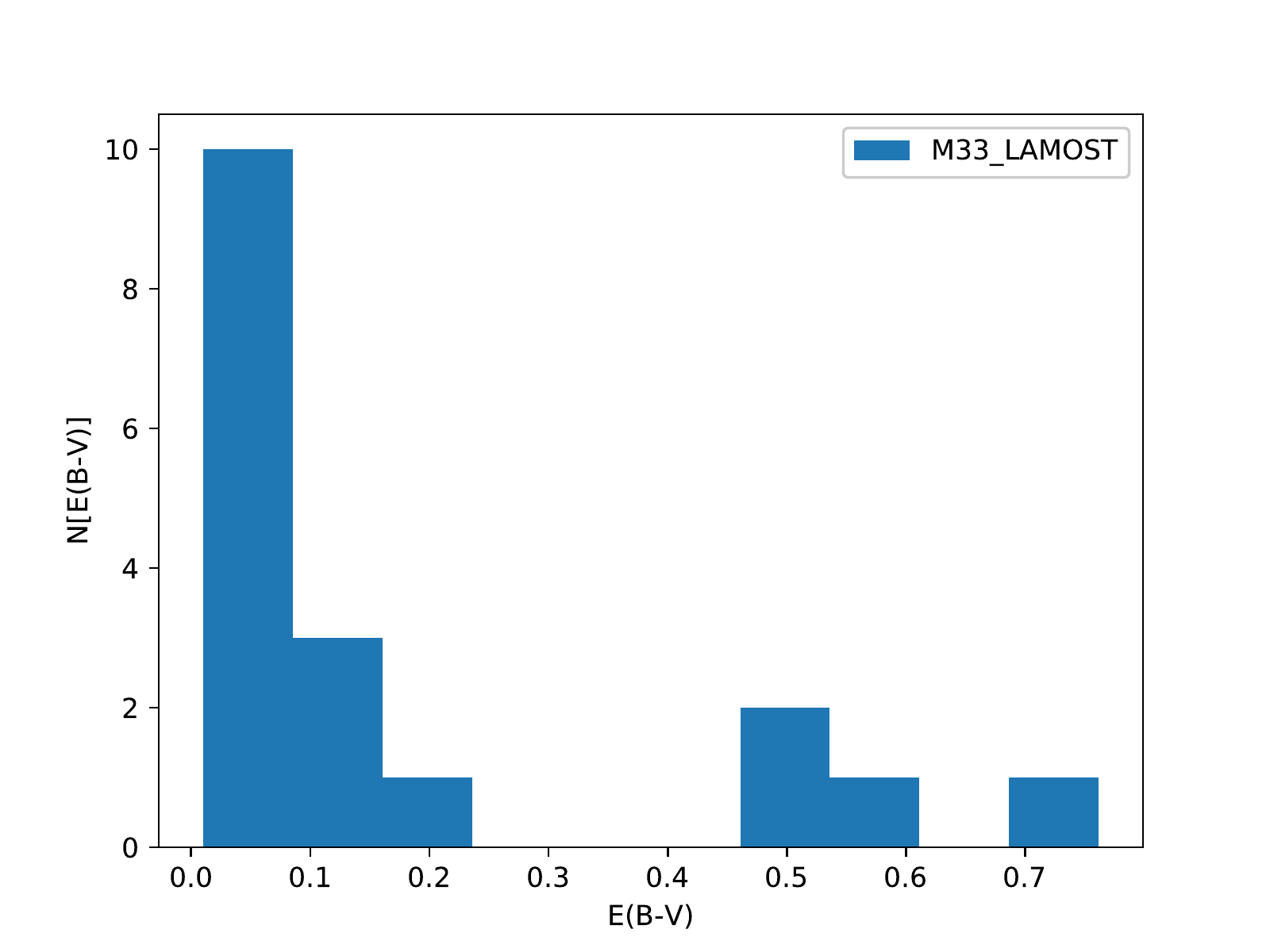}
\includegraphics[width=0.50\textwidth]{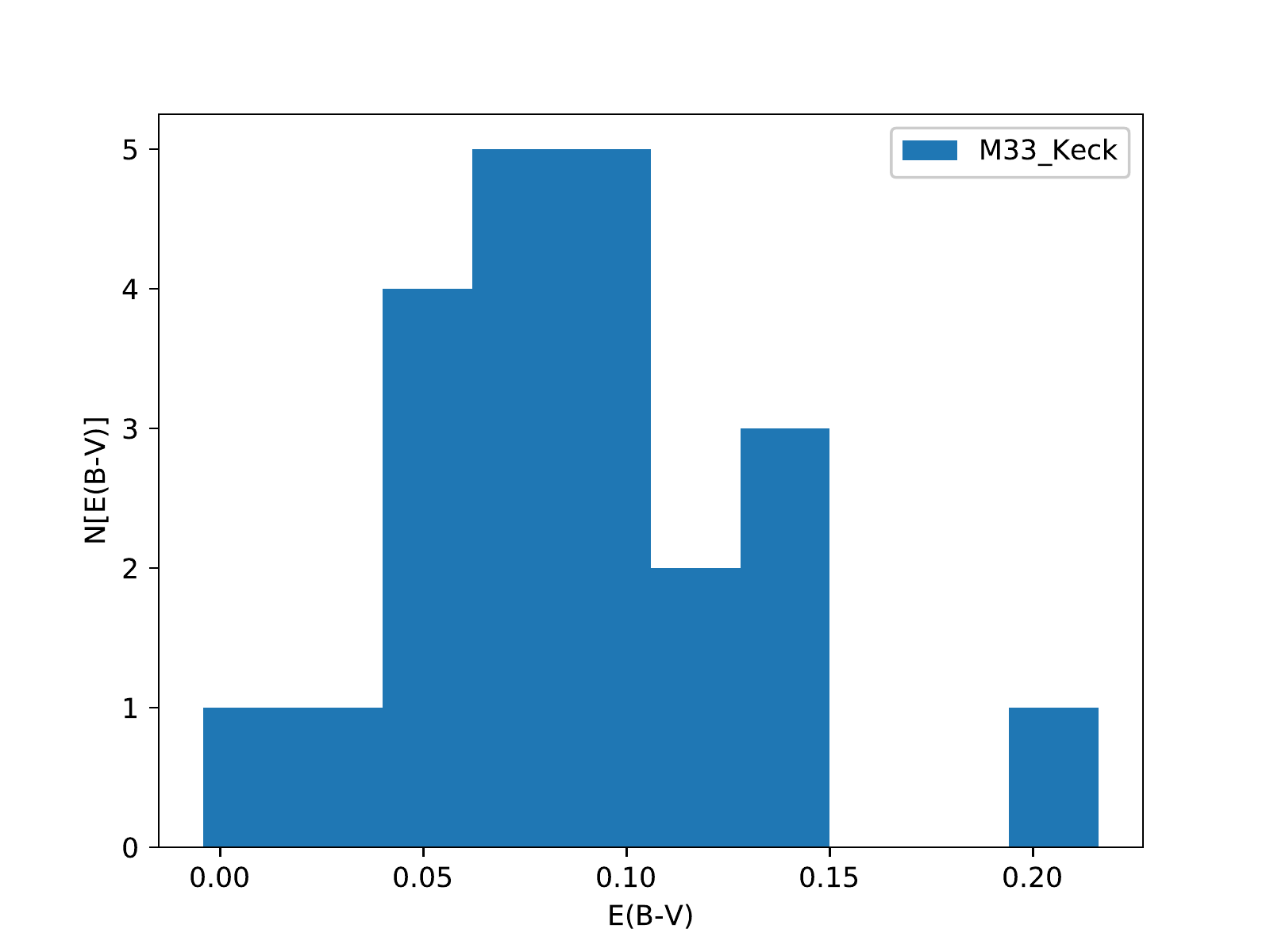}
\caption{Distribution of BSG reddening $E(B-V)$ for for the M 31 and M 33 samples.}
\label{sect5:EBV}
\end{figure}

For M 33  there is a wide reddening range from 0.01 to 0.8 mag with an average of $<E(B-V)>=0.14$ mag (see the middle and bottom plans of Figure \ref{sect5:EBV}). This is slightly higher than the average of 0.08 mag from the sample of BSGs in \cite{2009ApJ...704.1120U}, because of four extreme cases with  $E(B-V) \ge 0.47$ mag. Our result is smaller than the value of 0.20 mag adopted in the $\it{HST}$ distance scale Key Project study of Cepheids by \cite{1991ApJ...372..455F} and the average value ($<E(B-V)>=0.22$ mag) from the study of H {\scriptsize $\mathrm{II}$} regions in \cite{2007A&A...470..865M}. Using published reddening values from a large sample of H {\scriptsize $\mathrm{II}$} regions from \cite{2008ApJ...675.1213R}, \cite{2009ApJ...704.1120U} produced $E(B-V)$ as a function of angular galactic distance and found a large scatter in reddening at all distance \citep[see Figure 9 in][]{2009ApJ...704.1120U}. The more recent work on 413 star-forming (or H {\scriptsize $\mathrm{II}$}) regions in M33 also indicates a large scatter in reddening and shows several extreme cases with $E(B-V) > 0.8$ mag in the outer disk \citep{2017ApJ...842...97L}. However, the average value of $<E(B-V)>=0.13$ mag is in agreement to what is calculated from our BSG samples in M 33.

\subsection{Stellar Properties}\label{ste-pro}
\begin{figure}[ht!]
\centering
\includegraphics [width=0.45\textwidth]{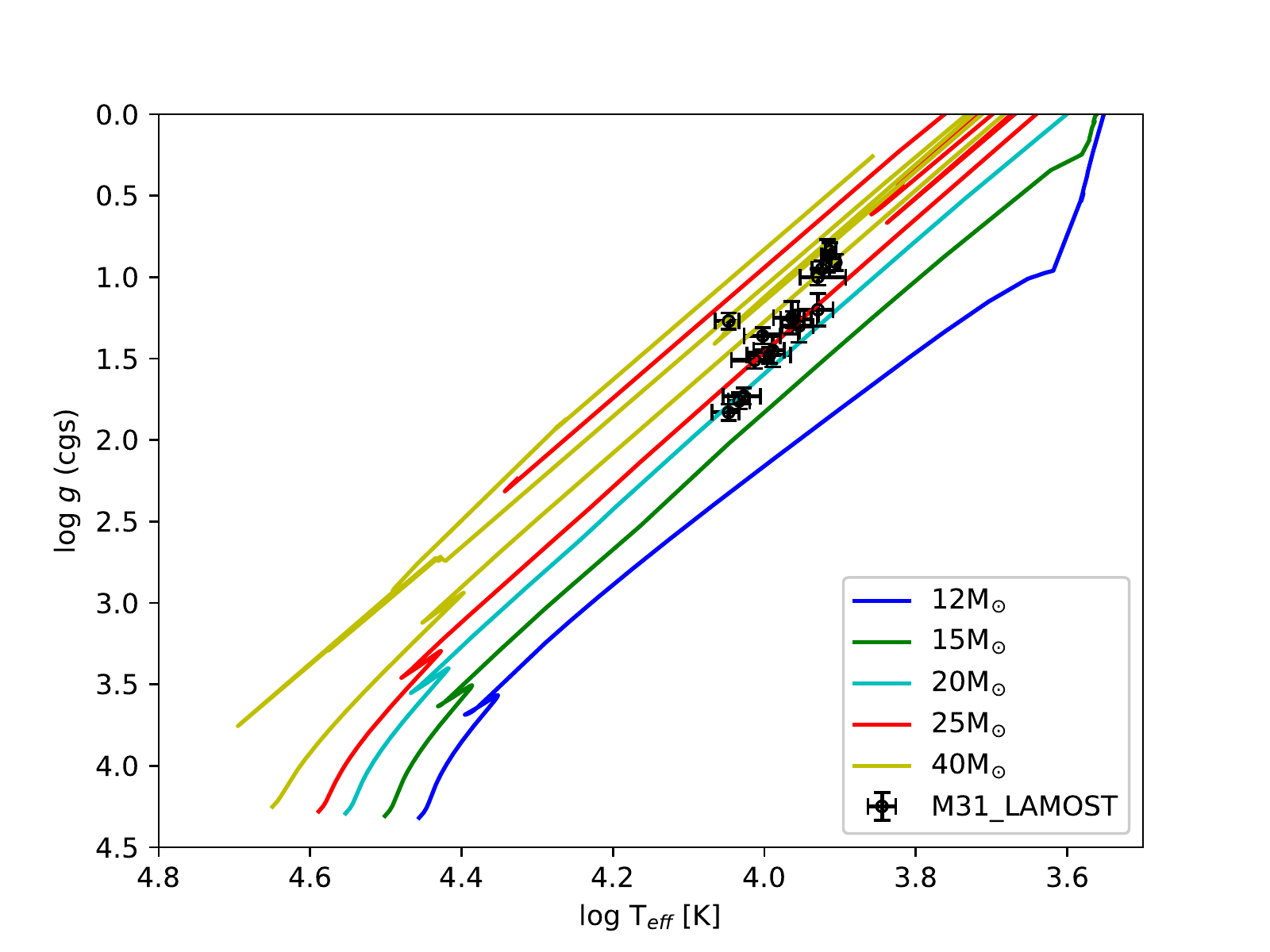}
\includegraphics [width=0.45\textwidth]{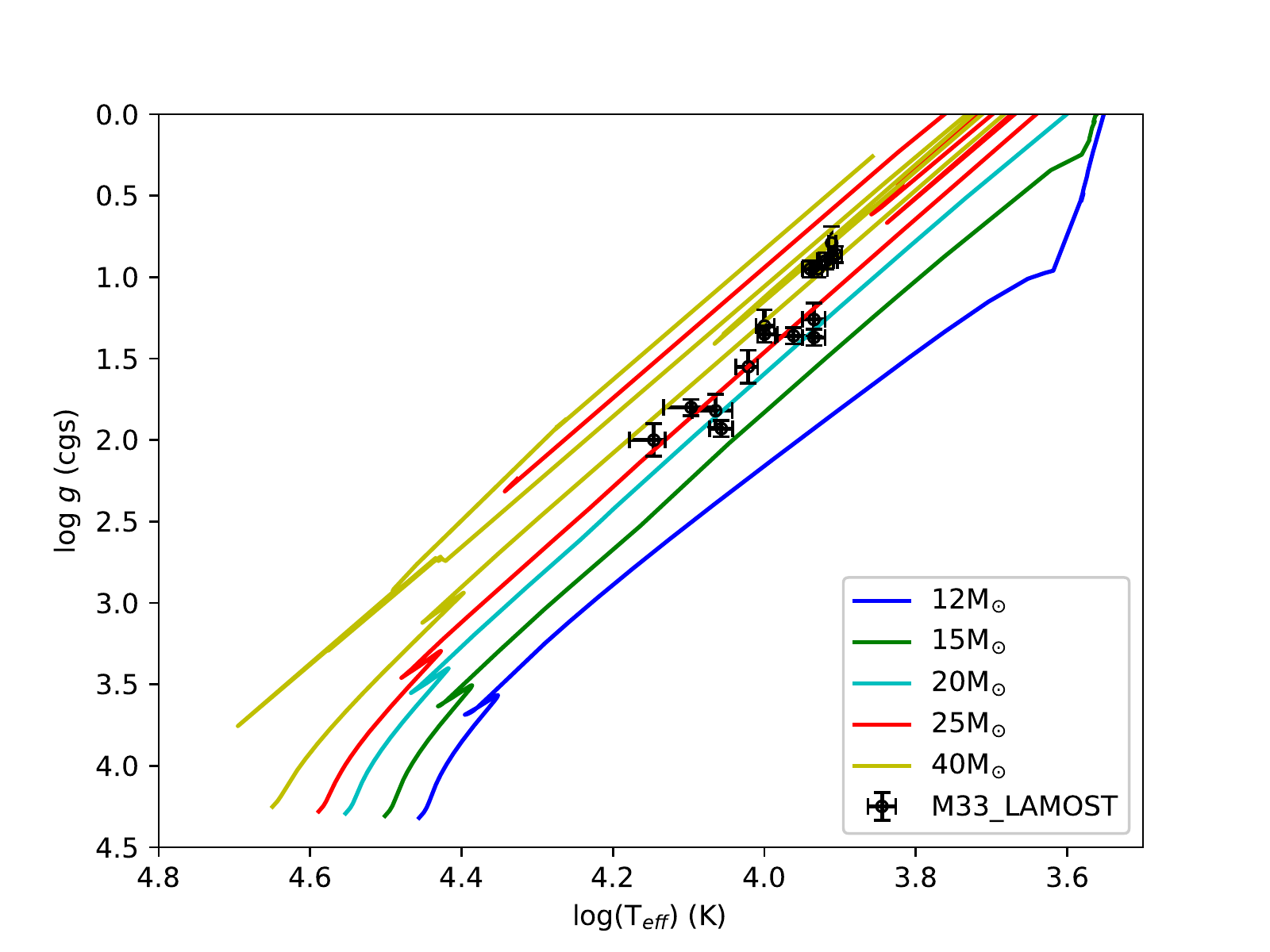}
\includegraphics [width=0.45\textwidth]{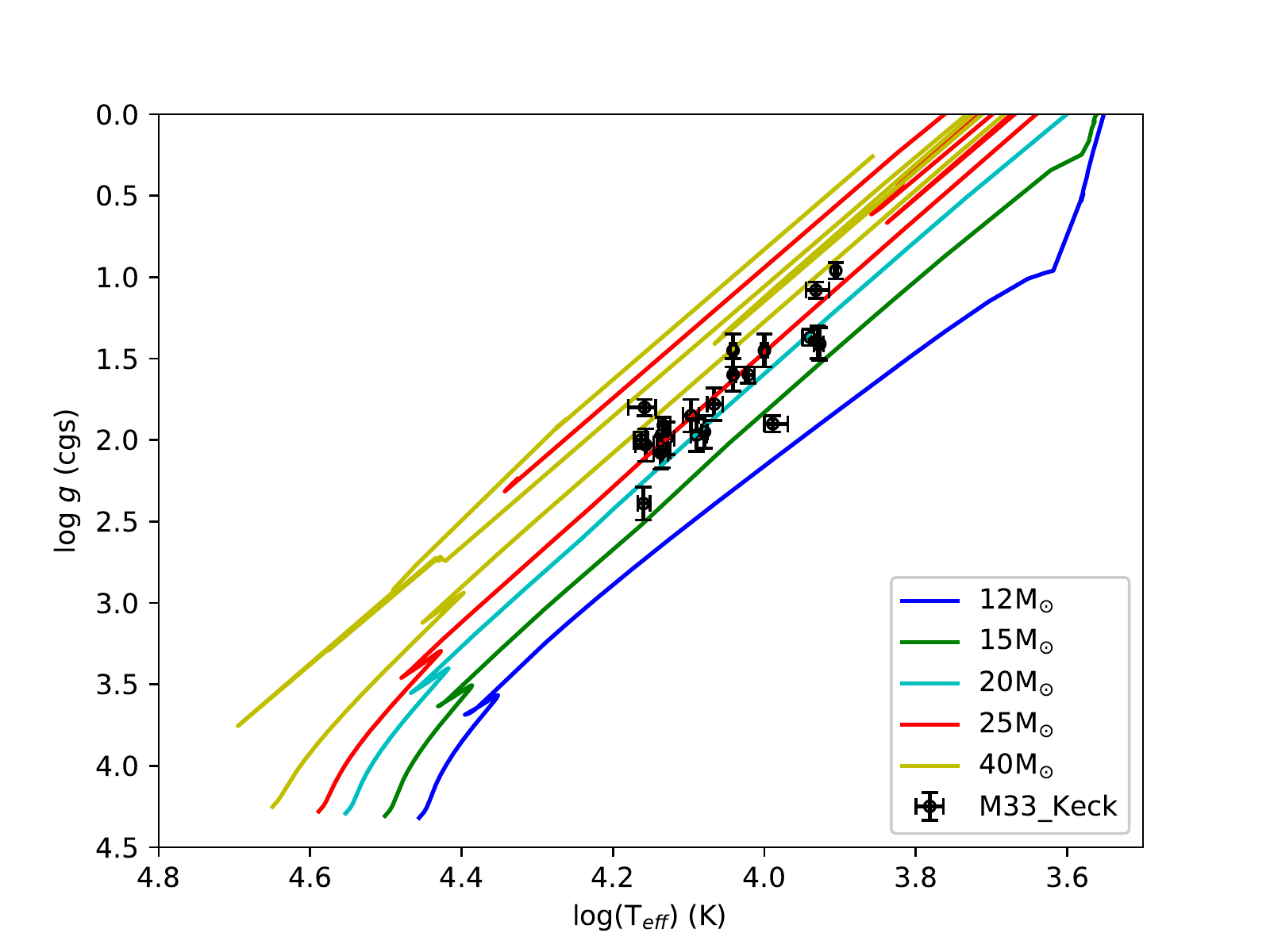}
\caption{Gravities and temperatures of the observed BSGs compared with evolution tracks (with 12, 15, 20, 25, and 40 $M_{\odot}$, respectively) including the effects of rotational mixing \citep{2012A&A...537A.146E}). Top: M 31 BSG sample. Middle: M 33 members from LAMOST survey. Bottom: M 33 members with Keck DEIMOS spectra.\label{sect5:T-G}}
\end{figure}

\begin{figure}[ht!]
\centering
\includegraphics [width=0.5\textwidth]{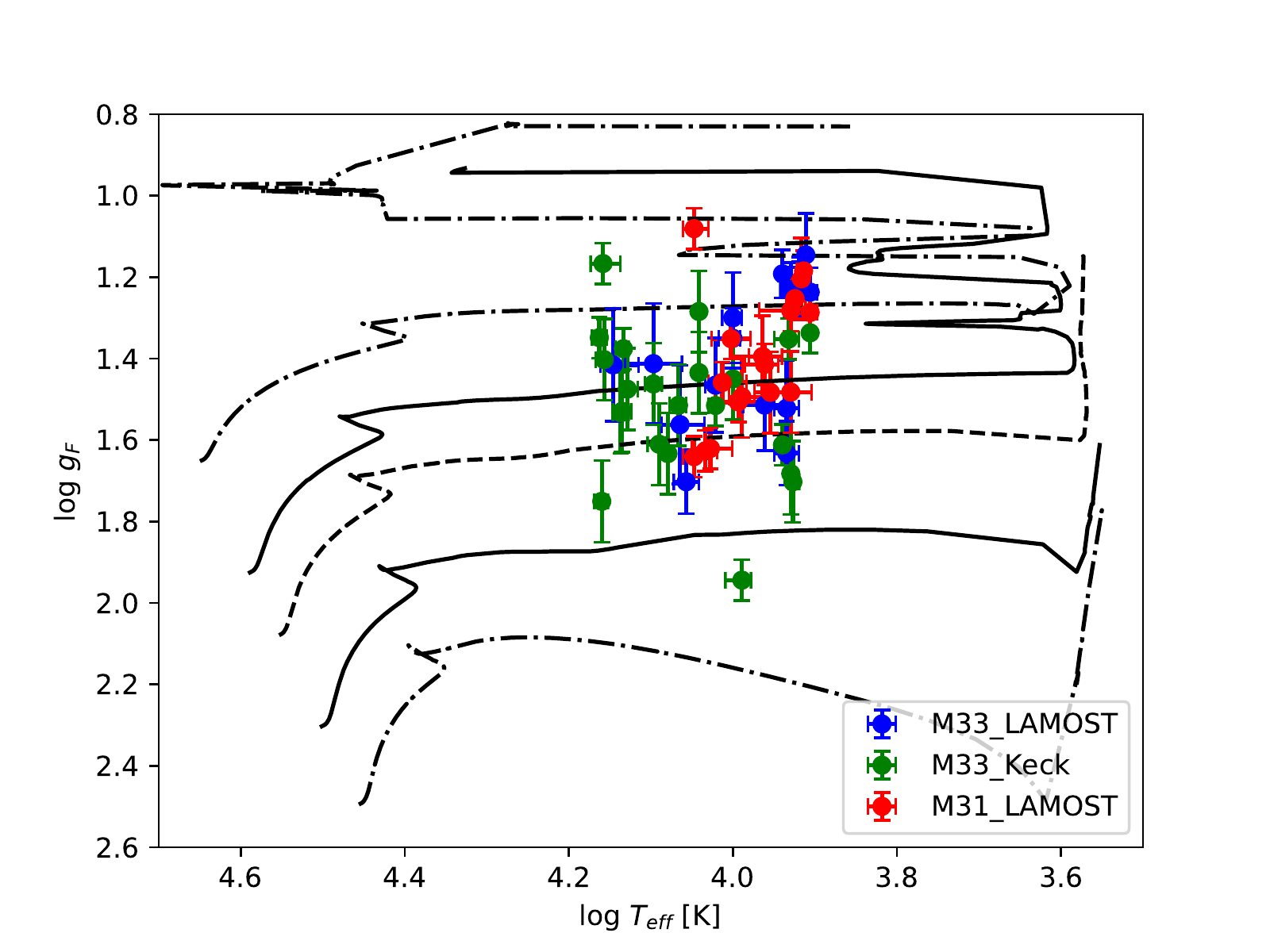}
\caption{Spectroscopic Hertzsprung-Russell diagram of BSGs of this study compared with evolutionary tracks from \cite{2012A&A...537A.146E}. Red: M 31 members, blue: the M 33 members observed with LAMOST, green: the M 33 members observed with Keck. The evolutionary tracks are calculated for initial main-sequence masses of (from the bottom of the figure to top) 12, 15, 20, 25, and 40 $M_{\odot}$. 
\label{sect5:sHRD}}
\end{figure}

With the stellar effective temperatures and gravities obtained from the spectroscopic analysis, we can discuss stellar properties and evolutionary status of the BSGs. Figure \ref{sect5:T-G} shows the locations  in the ($T_{\rm{eff}}$, log $g$) diagram compared with evolutionary tracks. As the diagram is independent of the distances, it allows to investigate the properties of the target stars without being effected by distance uncertainties. It clearly shows that the BSGs originally had masses of around 25 $M_{\odot}$ and have now evolved away from the main sequence. Although the uncertainties in log $g$ do not allow for a clear conclusion, it looks like that there are several stars heavier than 40 $M_{\odot}$, and one target seems to be less massive than 15 $M_{\odot}$. The evolutionary status of the objects can be assessed from Figure \ref{sect5:sHRD}, which shows flux-weighted gravities ($g_{\rm{F}}$ = $g$/$T_{\rm{eff}}^4$, $T_{\rm{eff}}$ in units of $10^4$ K) against $T_{\rm{eff}}$ and is morphologically similar to the classical Hertzsprung-Russell diagram. This spectroscopic Hertzsprung-Russell diagram (sHRD) has recently been introduced and discussed in detail by \cite{2014A&A...564A..52L}. It provides the advantage to study stellar evolution independent of distance and clearly indicates that most of stars are in a mass range from 15 to 40 $M_{\odot}$.

Assuming a distance modulus determined via the FGLR method (see below), we can  determine stellar masses, radii, bolometric magnitudes, and luminosities with the knowledge of reddening and extinction. As demonstrated in \cite{2008ApJ...681..269K}, there are two ways to determine stellar masses. We can use the stellar gravities together with the radii to directly calculate spectroscopic masses. Alternatively, evolutionary masses can be determined by comparing the location of BSGs on the HR diagram with evolutionary tracks. Both mass estimates are given in Table \ref{M31_sp} and \ref{M33_sp} for M 31 and M 33 samples, respectively. The uncertainties of the spectroscopic masses are larger than those of the evolutionary masses because of the relatively large uncertainties of log $g$ compared with the uncertainties of $T_{\rm{eff}}$ and photometric data. On the other hand, possible systematic uncertainties of the evolutionary tracks are not included in the evolutionary masses. 

\begin{figure}[ht!]
\centering
\includegraphics[width=0.50\textwidth]{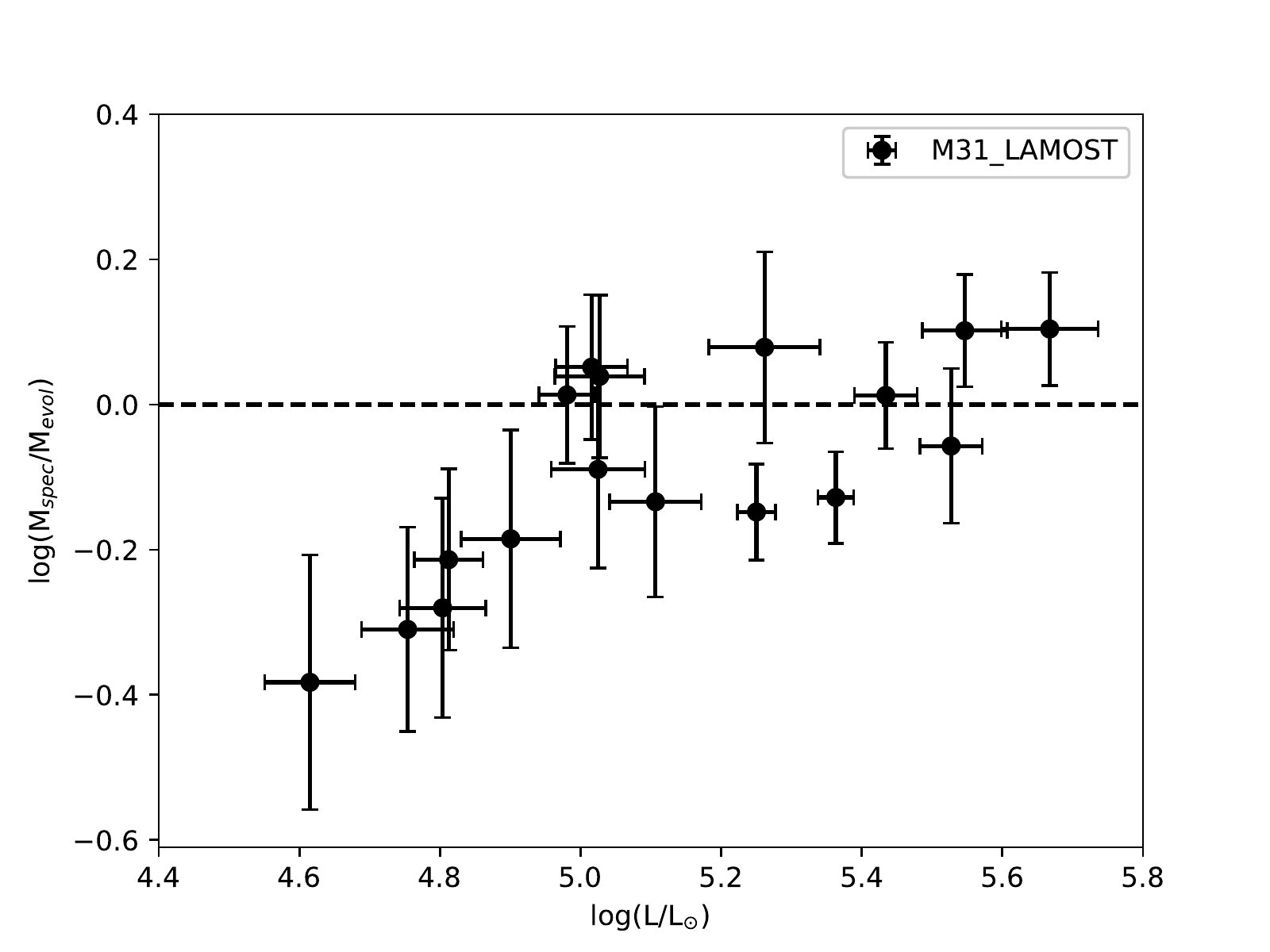}
\includegraphics[width=0.50\textwidth]{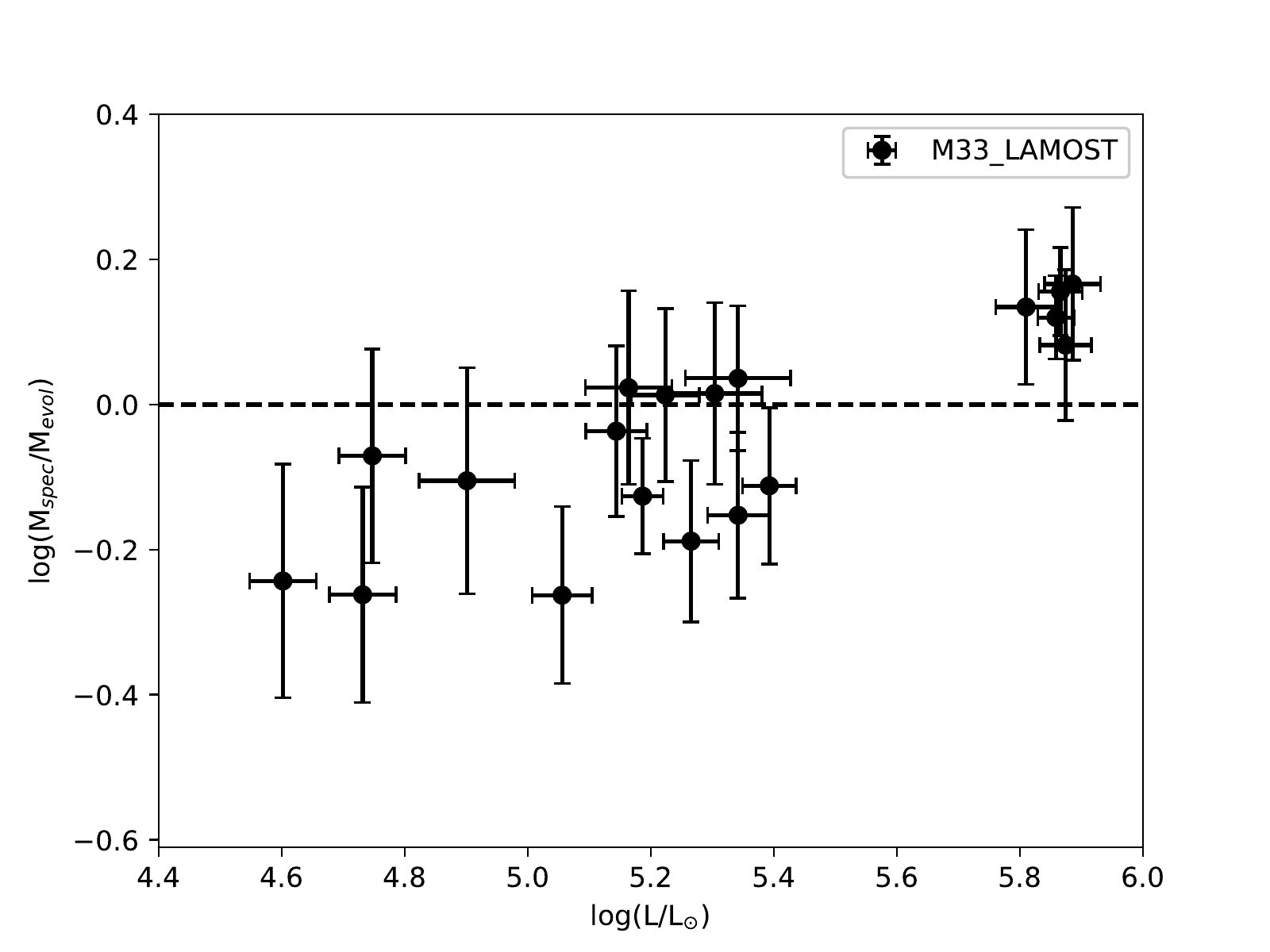}
\includegraphics[width=0.50\textwidth]{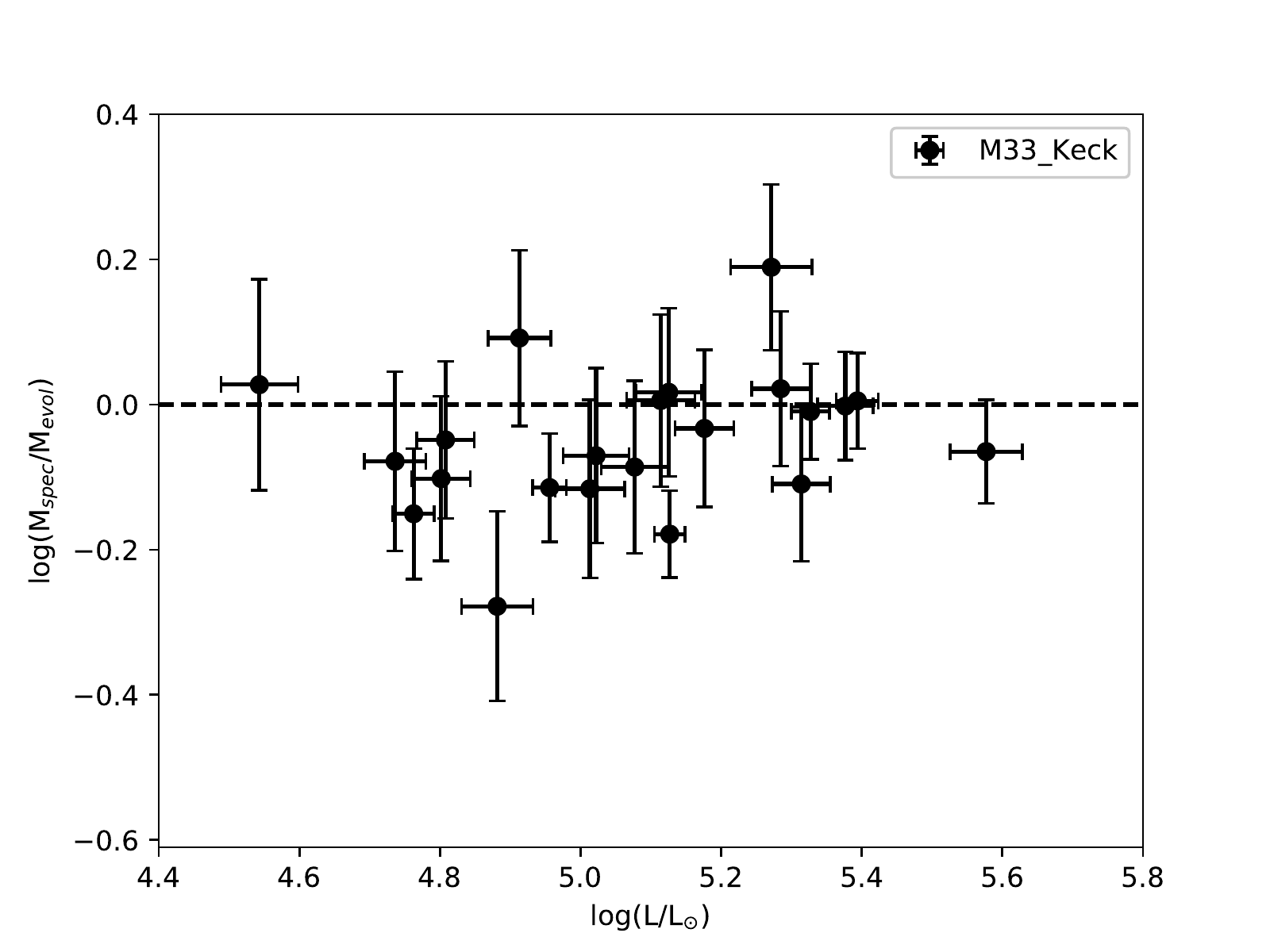}
\caption{Logarithmic ratio of spectroscopic to evolutionary masses as a function of luminosity. Top: M 31 BSG sample. Middle: M 33 members (LAMOST). Bottom: M 33 members (Keck).}
\label{sect5:M-L}
\end{figure}

A comparison of two sets of stellar masses is shown in Figure \ref{sect5:M-L}. Our results are similar to what has been found in  previous extragalactic studies of BSGs \citep[e.g.][]{2008ApJ...681..269K, 2009ApJ...704.1120U, 2014ApJ...788...56K, 2014ApJ...785..151H}. Spectroscopic masses seem to be smaller, in particular at lower luminosities. For the two M 33 samples the logarithmic difference is smaller by 0.05 dex. For M 31 the difference is 0.07 dex if we exclude the one extreme target J004005.02+403242.2 with log$(M_{\rm{spec}}/M_{\rm{evol}}) \simeq -0.38$. As discussed in \cite{2008ApJ...681..269K} and \cite{2009ApJ...704.1120U} for similar cases the spectroscopic  mass of this object may have been affected by close binary or blue loop evolution with strong mass-loss. The general trend that the spectroscopic masses are smaller than their evolutionary counterparts at the lower luminosity has already been found in the previous BSG studies by \cite{2008ApJ...681..269K, 2012ApJ...747...15K}. It is related to the "mass discrepancy" problem of massive stars discovered first by \cite{1992A&A...261..209H} (see also \citealt{2009msfp.book..126K}) and indicates a potential systematic uncertainty in the spectral diagnostics or stellar evolution theory or both. While there is a rich literature on this subject, the problem remains unsolved.

\section{Metallicity and Metallicity gradient} \label{sec:MG}
\begin{figure}
\centering
\includegraphics[width=0.50\textwidth]{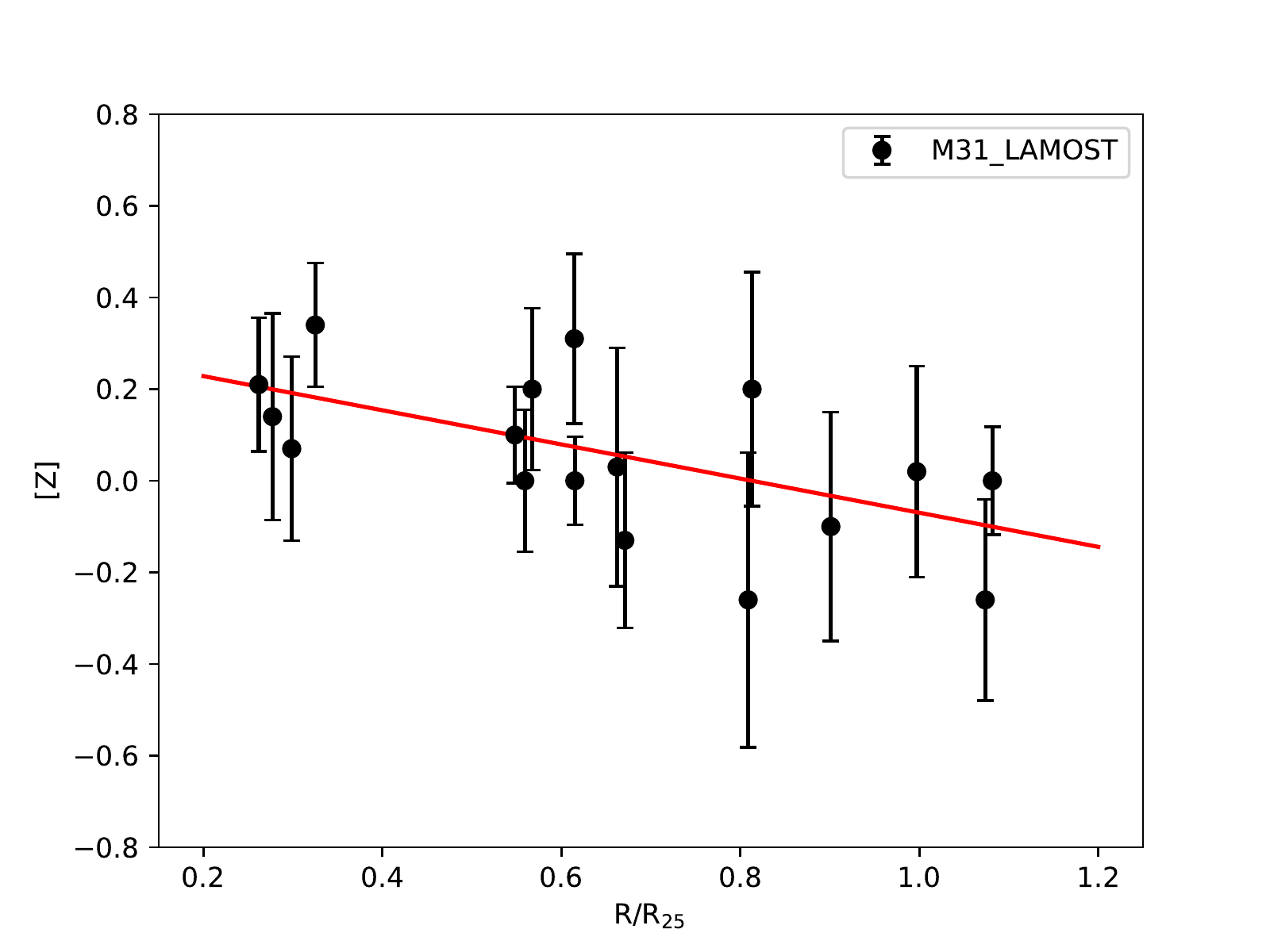}
\includegraphics[width=0.50\textwidth]{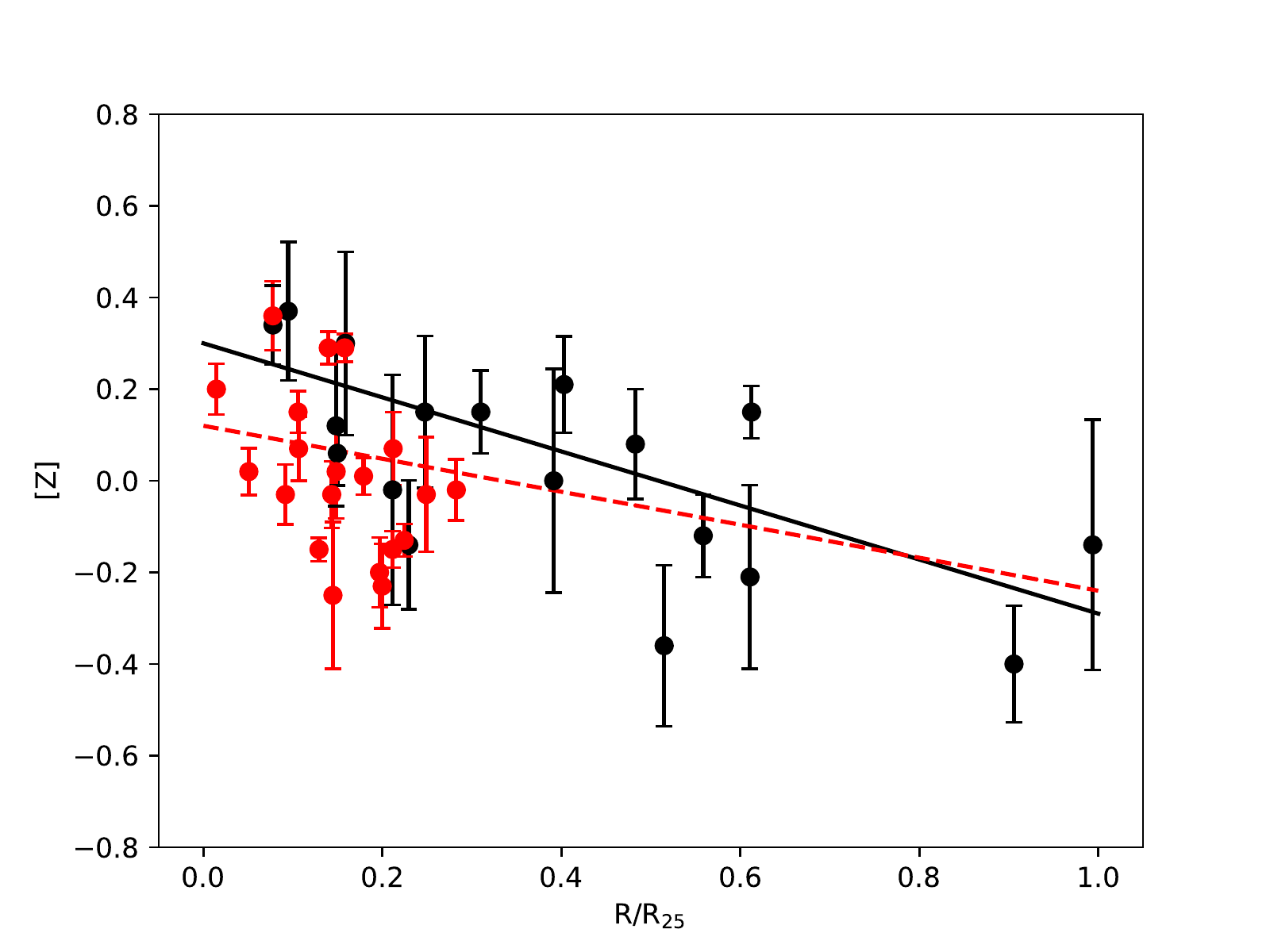}
\caption{Metallicity of the blue supergiants in M 31 (top) and M 33 (bottom) as a function of galactocentric distance. In M 31 the linear regression is shown in red. For M 33 metallicities from the LAMOST sample and the corresponding regression are plotted in black. The metallicities of the Keck sample are given in red. The regression for the combined LAMOST/Keck samples is represented by the red dashed line.  
}
\label{Z-grad}
\end{figure}

The metallicities of all targets together with their galactocentric distances are given in Table \ref{M31_sp} and \ref{M33_sp}. Figure \ref{Z-grad}  displays metallicities as a function of the dimensionless angular galactocentric distance $R/R_{25}$ (the galactocentric distance normalized to the isophotal radius of the disc, $R_{25}$) for both galaxies. A simple linear regression (accounting for errors in metallicities) of the form 
\begin{equation}
[Z] = [Z]_{0} + [Z]_{1}(R/R_{25}),
\label{eq:Z}
\end{equation}
approximates the observed spatial distribution in terms of an average metallicity gradient. 

\subsection{Metallicity and Metallicity gradient in M 31}\label{sec:M31-M}
For M 31 sample we obtain $[Z]_{0}=0.30\pm0.09$ dex for the central metallicity and $[Z]_{1} = -0.37\pm0.13$ dex/$R_{25}$ for the angular gradient. This value is in good agreement with the typical galactic 'benchmark' metallicity gradient  $[Z]_{1} = -0.42\pm0.16$ dex/$R_{25}$ found by  \cite{2015MNRAS.448.2030H} from a large sample of local star forming galaxies. The angular gradient is equivalent to $-0.018\pm0.006$ dex kpc$^{-1}$ in distance.

\begin{figure}
\centering
\includegraphics[width=0.50\textwidth]{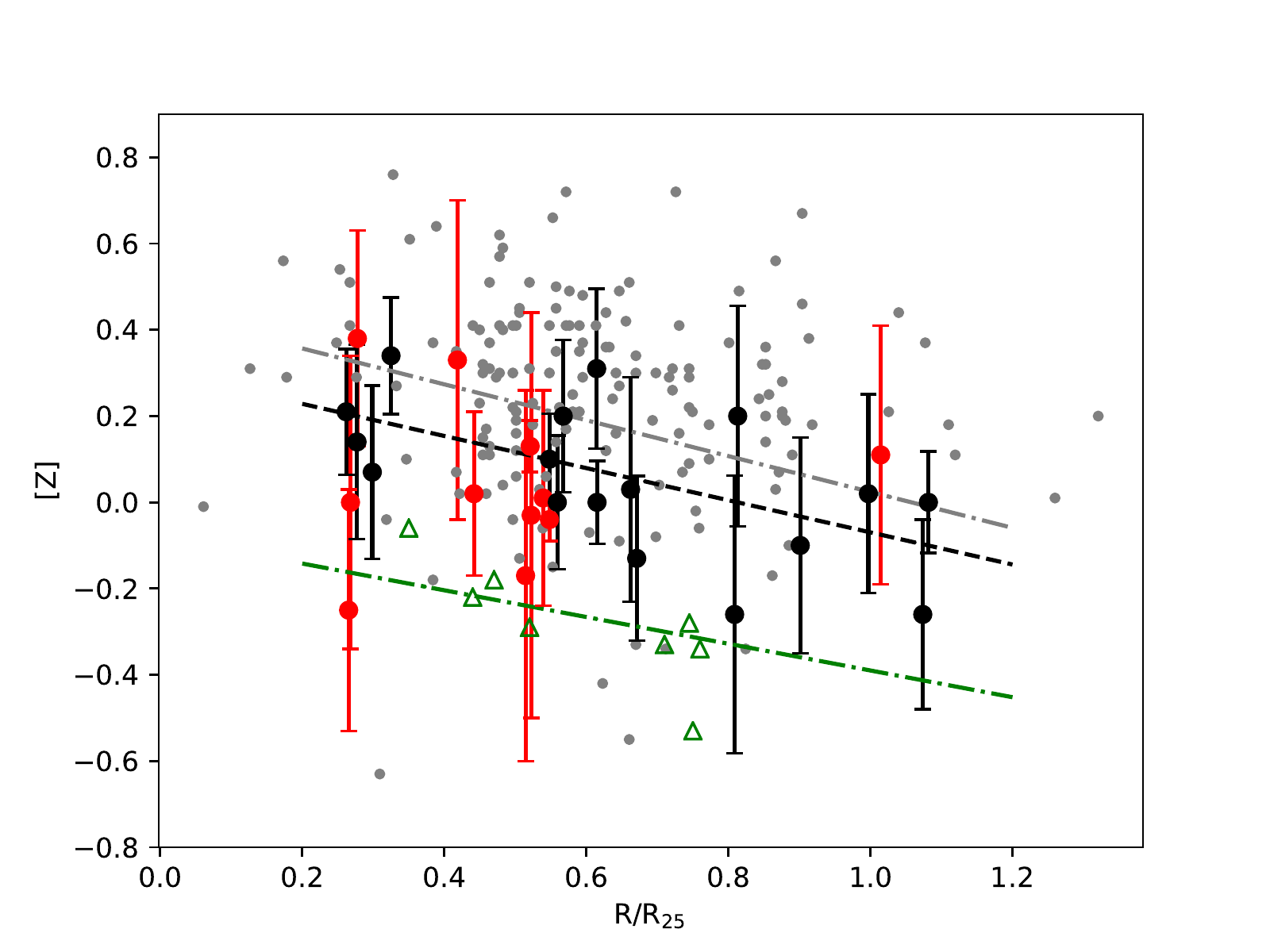}
\caption{M 31: Metallicity of the blue supergiants and H {\scriptsize $\mathrm{II}$} regions as a function of galactocentric distance. The BSG of this study  are plotted in black with the same symbols as Figure \ref{Z-grad}, while BSG metallicities from previous work (see text)  are shown as the red full circles. Logarithmic oxygen abundances of H {\scriptsize $\mathrm{II}$} regions in units of the solar value as published by \cite{2012ApJ...758..133S} and \cite{2020MNRAS.491.2137E} are plotted as the gray points and green open triangles, respectively. The three linear regressions corresponding to our sample and two H {\scriptsize $\mathrm{II}$} region studies are plotted in dashed black and dashed-dotted gray and green, respectively.}
\label{Z-grad-ref-m31}
\end{figure}

In Figure \ref{Z-grad-ref-m31} we compare our results with previous work on BSG \citep{2000ApJ...541..610V,2001MNRAS.325..257S,2002A&A...395..519T,2006astro.ph.11044P} and  H {\scriptsize $\mathrm{II}$} regions \citep{2012ApJ...758..133S, 2020MNRAS.491.2137E}. Our BSG results agree well with the previous BSG work. The metallicity gradients obtained from H {\scriptsize $\mathrm{II}$} regions are similar to the BSG gradient (--0.40 dex/$R_{25}$ and --0.31 dex/$R_{25}$, for Sanders et al. and Esteban et al., respectively), however the zero points differ significantly. This indicates a systematic effect in the H {\scriptsize $\mathrm{II}$} region metallicity zero point calibration.

\begin{figure}
\centering
\includegraphics[width=0.50\textwidth]{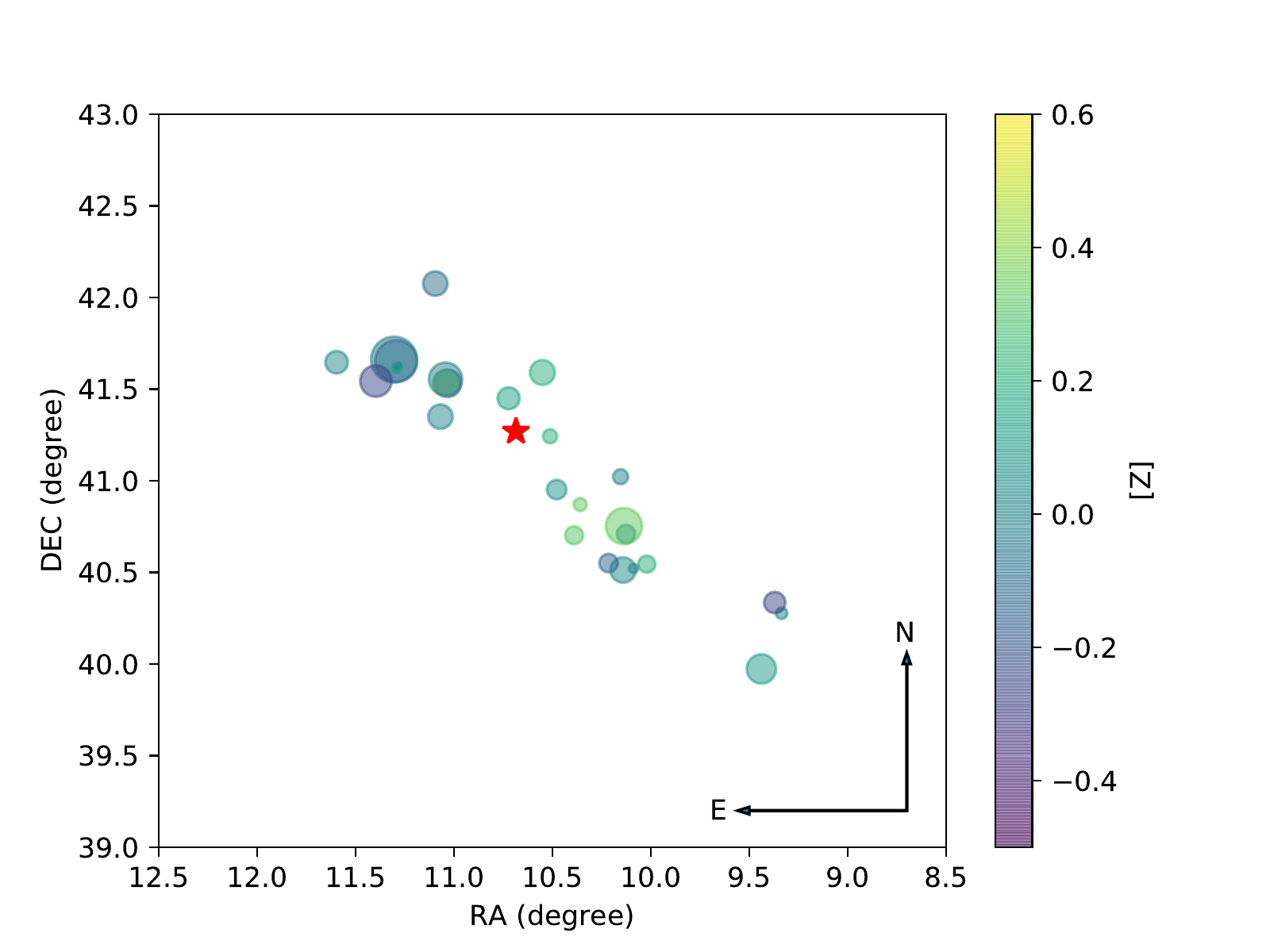}
\caption{The metallicity maps of M 31 disk based on our sample and BSGs from references \citep{2000ApJ...541..610V,2001MNRAS.325..257S,2002A&A...395..519T,2006astro.ph.11044P}. The colour-scale shows the metallicity as indicated in the label. A large symbol size represents a large uncertainty of metallicity, while a small symbol indicates a small uncertainty. The red star stands for the optical center of M 31.}
\label{Z-position-m31}
\end{figure}

Comparing these slopes with the abundance gradients from planetary nebulae (PNe), it is evident the young populations (supergiants and H {\scriptsize $\mathrm{II}$} regions) have a much steeper slope, while the old populations (PNe) have shallow gradients \citep[see details in][]{2012ApJ...753...12K} or even no significant trend at all \citep[see details in][]{2012ApJ...758..133S,2013ApJ...774....3B}. In addition, it is interesting to explore the 2-dimensional distribution of BSG metallicity (see Figure \ref{Z-position-m31}). Surprisingly, we find an off-center peak in the southeast direction. However, given the small number of BGS studied here this might be a selection effect and requires an extended investigation with an enlarged sample of objects. Such an extended study is also motivated by the result found for M 33 in Sect. \ref{sec:M33-M}, where the 2-D distribution reveals a clear off-centre peak.

\subsection{Metallicity and Metallicity gradient in M 33}\label{sec:M33-M}

For the M 33 members from LAMOST, we determine $[Z]_{0}=0.30\pm0.08$ dex and $[Z]_{1} = -0.54\pm0.16$ dex/$R_{25}$ (equivalent to $-0.06\pm0.02$ dex kpc$^{-1}$ in distance). The angular gradient is steeper than for M 31 but still with in the 1$\sigma$ range of the benchmark gradient found by  \cite{2015MNRAS.448.2030H}. Interestingly, the metallicities of the stars in our Keck sample appear to be systematically lower than those of the LAMOST sample. A regression of the combined LAMOST/Keck samples yields  $[Z]_{0}=0.11\pm0.04$ dex and $[Z]_{1} = -0.36\pm0.16$ dex/$R_{25}$ with a lower central metallicity and a shallower gradient. The BSG study by \cite{2009ApJ...704.1120U} obtained a central metallicity   $[Z]_{0}=0.09\pm0.04$ dex in agreement with our combined sample but derived a much steeper gradient $[Z]_{1} = -0.73\pm0.09$ dex/$R_{25}$. The disagreement is puzzling but we will try to give an explanation below. 

We also compare with metallicity information from H {\scriptsize $\mathrm{II}$} regions. We combine two sets of O abundances from \cite{2007A&A...470..865M,2010A&A...512A..63M} together with two sets of Ne and N abundances from \cite{2008MNRAS.387...45R} and \cite{2016MNRAS.458.1866T}, respectively, and derive a slope of $-0.57\pm0.07$ dex/$R_{25}$ and a central metallicity of [Z]$_{0}=-0.08\pm0.03$ dex.

\begin{figure}
\centering
\includegraphics[width=0.50\textwidth]{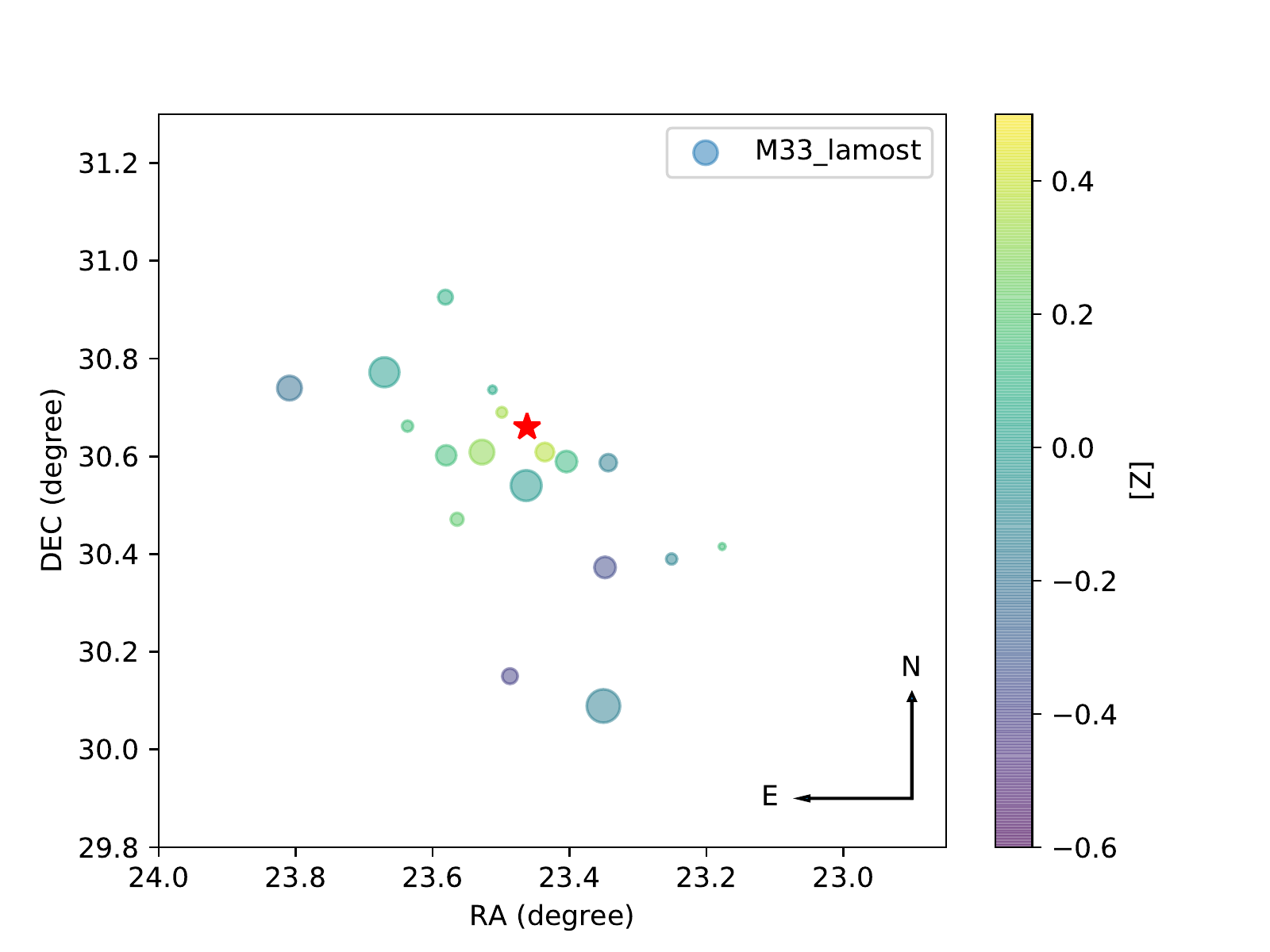}
\includegraphics[width=0.50\textwidth]{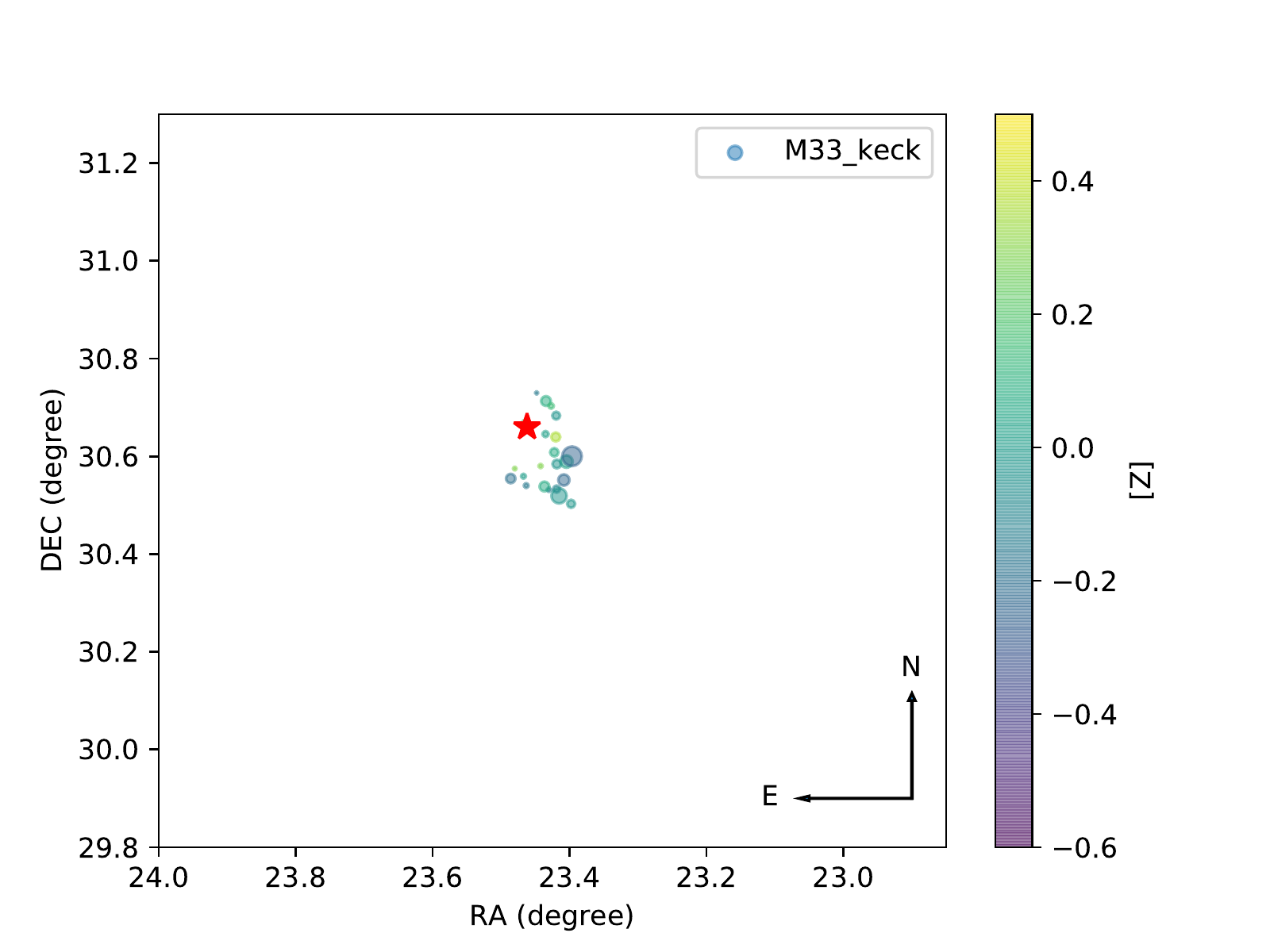}
\includegraphics[width=0.50\textwidth]{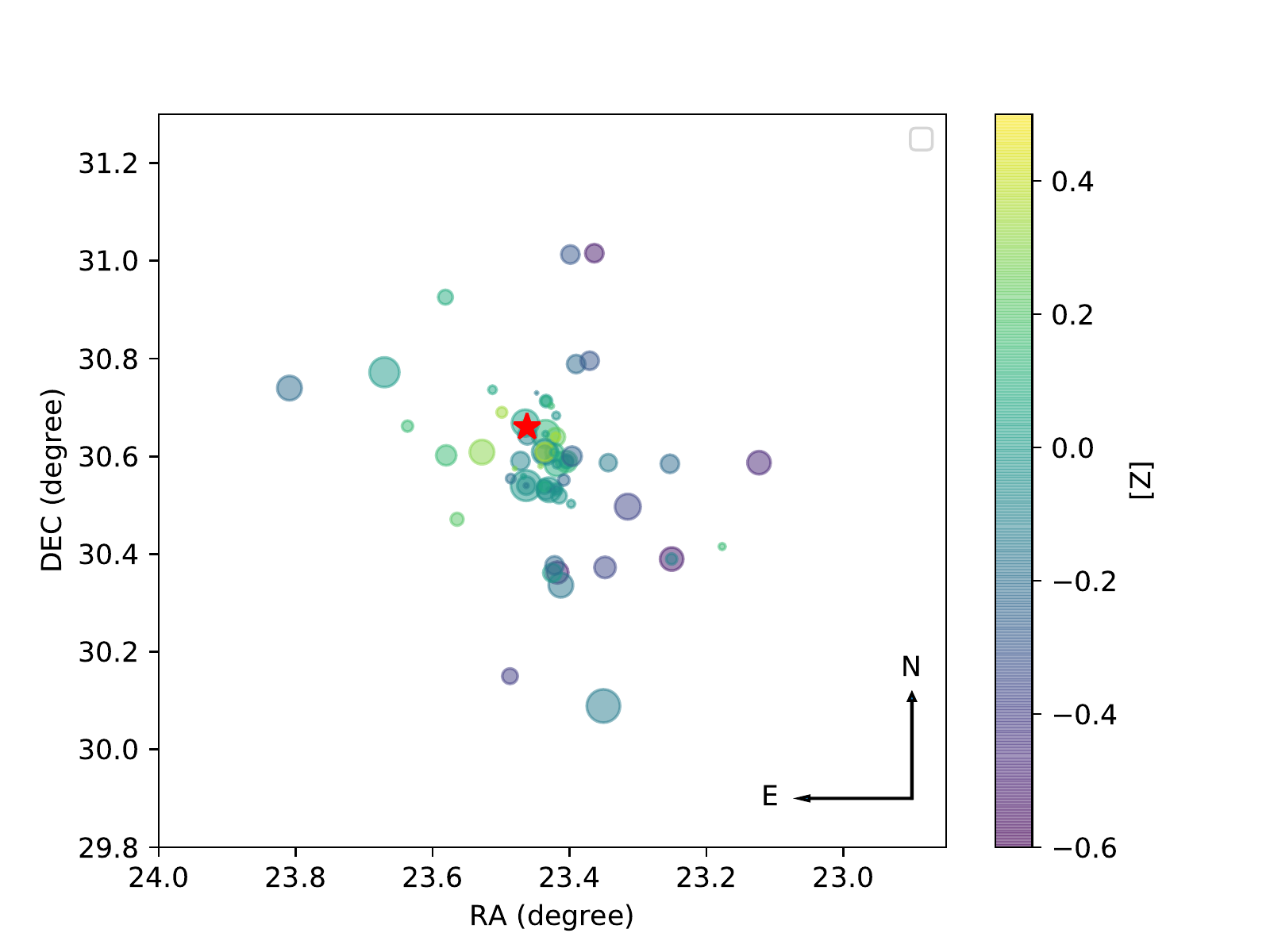}
\caption{The metallicity maps of M 33 disk based on our samples and BSGs from \cite{2009ApJ...704.1120U}. The colour-scale and symbols are the same as in Figure \ref{Z-position-m31}. Top: M 33 members from LAMOST survey. Middle: M 33 members observed by the Keck telescope. Bottom: combining two top samples and BSGs from \cite{2009ApJ...704.1120U}.}
\label{Z-position}
\end{figure}

The disagreement between the different results is puzzling. Motivated by the results by \cite{2007A&A...470..865M},  who found a very steep H {\scriptsize $\mathrm{II}$} region metallicity slope in the central 3 kpc region of M 33, and the discovery of an off-center metallicity peak by \cite{2010A&A...512A..63M} we investigate the 2-dimensional distribution of BSG metallicity in Figure \ref{Z-position}. We find that the metallicity distribution of all BSGs, including the targets from \cite{2009ApJ...704.1120U}, in our investigation is not axially symmetric. There is a clear trend that the metallicity is much poorer in the eastern part than the metallicity at the same distance of the western part of M 33 (see the bottom panel of Figure \ref{Z-position}). Unlike \cite{2010A&A...512A..63M} who discovered that the highest metallicity lie in the southern arm at 1-2 kpc from the center, we find that the location of the BSG off-center metallicity peak in our study might be located in the northwest direction. This clearly warrants an extended investigation.

\subsection{A comparison with the Galactic metallicity gradient}
It is interesting to compare the metallicity gradients of M 31 and M 33 with our Milky Way. Extensive investigations have been carried out for the Milky Way using different metallicity tracers such as OB stars, Cepheids, H {\scriptsize $\mathrm{II}$} regions and PNe. For H {\scriptsize $\mathrm{II}$} regions, most studies obtain remarkably consistent slopes of the oxygen gradient (between --0.04 dex kpc$^{-1}$ and --0.06 dex kpc$^{-1}$), even taking into account different methodologies \citep[e.g.][]{2017A&A...597A..84F, 2018MNRAS.478.2315E}. Slightly steeper gradients (between --0.06  dex kpc$^{-1}$ and --0.07  dex kpc$^{-1}$) were obtained from OB stars \citep[see][and references therein]{2019A&A...625A.120B}. Comprehensive work with Cepheids resulted in a gradient of --0.06  dex kpc$^{-1}$ \citep{2014A&A...566A..37G}. This means that on a linear kpc scale the metallicity gradient of our Galaxy is substantially larger (factor 2--4) than that of the M 31 disk (--0.018 dex kpc$^{-1}$). However, if we express the metallicity gradients in terms of the disk effective radius ($R_{\rm{e}}=1.678 \times R_{\rm{d}}$ valid under the assumption of an exponential disc \citep[see][]{2014A&A...563A..49S}), then the scaled gradient of the MW disk is similar to that of M 31 and M 33 disks (see Table \ref{tab:comp-gd}). This is in agreement with \cite{2020MNRAS.496.1051A} and \cite{2015MNRAS.448.2030H}, who found that most nearby galaxies have very similar scaled gradients. Comparing the scalelength of $R_{25}$ and $R_{\rm{d}}$, $R_{\rm{e}}$ has been considered as the best to normalize the metallicity gradients \citep{1989epg..conf..377D}. It should be mentioned that for M 33 the gradient of --0.36 dex/$R_{25}$ from the combined LAMOST/Keck samples is used rather than the value from only LAMOST sample in Section \ref{sec:M33-M}.

Furthermore, most of the determinations of the Galactic oxygen gradient obtained from PNe have typical values in the range between --0.02 and --0.04 dex kpc$^{-1}$ \citep{2013RMxAA..49..333M, 2019MNRAS.482.3071M}, which are flatter than the values from young tracers, such as H {\scriptsize $\mathrm{II}$} regions and OB stars. The same phenomenon is found in M 31 disk as discussed in Section \ref{sec:M31-M}. 
Compared with the gradients from younger tracers, the studies of PNe indicate that the radial gradient of metallicity was flatter in the past than in the present time. This is consistent with the prediction that a gas pre-enriched disc naturally develops an initial flat metallicity gradient that becomes steeper with time \citep{2001ApJ...554.1044C}.

\begin{table}[htp]
\centering
\caption{Radial metallicity gradients in galaxies.}
\begin{tabular}{rcccc}
\hline
Galaxy &Metallicity gradient &Ref. & $R_{\rm{e}}$ & Ref.\\
            & dex/$R_{\rm{e}}$  & & kpc & \\
\hline
The MW &--0.15 $\sim$ --0.25 & 1, 2, 3, 4 & 3.61 & 5\\
M 31 & --0.17 & this work & 8.89 & 6\\
M 33 & --0.15 & this work & 3.78 & 7\\
\hline
\end{tabular}
\label{tab:comp-gd}
\tablecomments{References. (1) \cite{2004ApJ...617.1115D}; (2) \cite{2018MNRAS.478.2315E}; (3) \cite{2019A&A...625A.120B}; (4) \cite{2020AJ....159..199D}; (5) \cite{2013ApJ...779..115B}; (6) \cite{2011ApJ...739...20C}; (7) \cite{1987AJ.....94..306K}}
\end{table}

\section{Distances\label{sect:dist}}
In this section, we employ the FGLR method to determine distances to M 31 and M 33. As discussed by \cite{2003ApJ...582L..83K,2008ApJ...681..269K} in detail, the FGLR is a tight correlation between the flux-weighted gravity ($g_{F} \equiv g/T^{4}_{\rm{eff}}$) and the absolute bolometric magnitude ($M_{\rm{bol}}$) of BA supergiants. The FGLR method is based on BSG spectroscopy and uses the determination of stellar gravities and effective temperatures which yield absolute magnitudes through this tight correlation and then distance moduli by comparison with derreddened apparent magnitudes. The crucial advantage of this method relative to the usual photometric methods using Cepheids or TRGB stars is that reddening and extinction are accurately determined because of the knowledge of spectroscopically determined temperatures and gravities. In addition, BSGs are much less affected by galactic stellar crowding.

The method has been applied to galaxies in the Local Group and beyond out to 7 Mpc distance  \citep{2012ApJ...747...15K, 2014ApJ...788...56K, 2016ApJ...829...70K, 2008ApJ...684..118U, 2009ApJ...704.1120U, 2014ApJ...785..151H, 2016ApJ...830...64B, 2018ApJ...860..130B}. Recently, \cite{2017AJ....154..102U} have introduced a new calibration of the FGLR based on the analysis of 90 BSG in the Large Magellanic Cloud (LMC). This new FGLR is divided into two parts at log $g_{F} = 1.30$ with two different slopes.

\begin{equation}
\rm{log~{\it{g}}_{F} \ge log {\it{g}}^{break}_{F} : {\it{M}}_{bol} = {\it{a}}(log~{\it{g}}_{F} - 1.5) + {\it{b}}}
\label{fglr-eq1}
\end{equation}
and
\begin{equation}
{\rm{log}}~g_{\rm{F}} \le {\rm{log}}~ g^{\rm{break}}_{\rm{F}} : M_{\rm{bol}} = a_{\rm{low}}({\rm{log}}~g_{\rm{F}} - {\rm{log}}~g^{\rm{break}}_{\rm{F}}) + b_{\rm{break}},
\label{fglr-eq2}
\end{equation}
with 
\begin{equation}
 b_{\rm{break}} = a({\rm{log}}~g^{\rm{break}}_{\rm{F}} - 1.5) + b
 \label{fglr-eq3}
\end{equation}
with log~$g^{\rm{break}}_{\rm{F}} =1.30$ dex, $a=3.20\pm0.08$, $b=-7.878\pm0.02$ mag, and $a_{\rm{low}} = 8.34\pm0.25$. Note that the value of b is marginally different from Urbaneja et al. accounting for the new 1 percent precision distance to the LMC measured by \cite{ 2019Natur.567..200P}. New FGLR distances using this improved calibration have been determined by \cite{2021ApJ...914...94S}.

\subsection{Distance to M 31}
\begin{figure}[ht]
\centering
\includegraphics[width=0.50\textwidth]{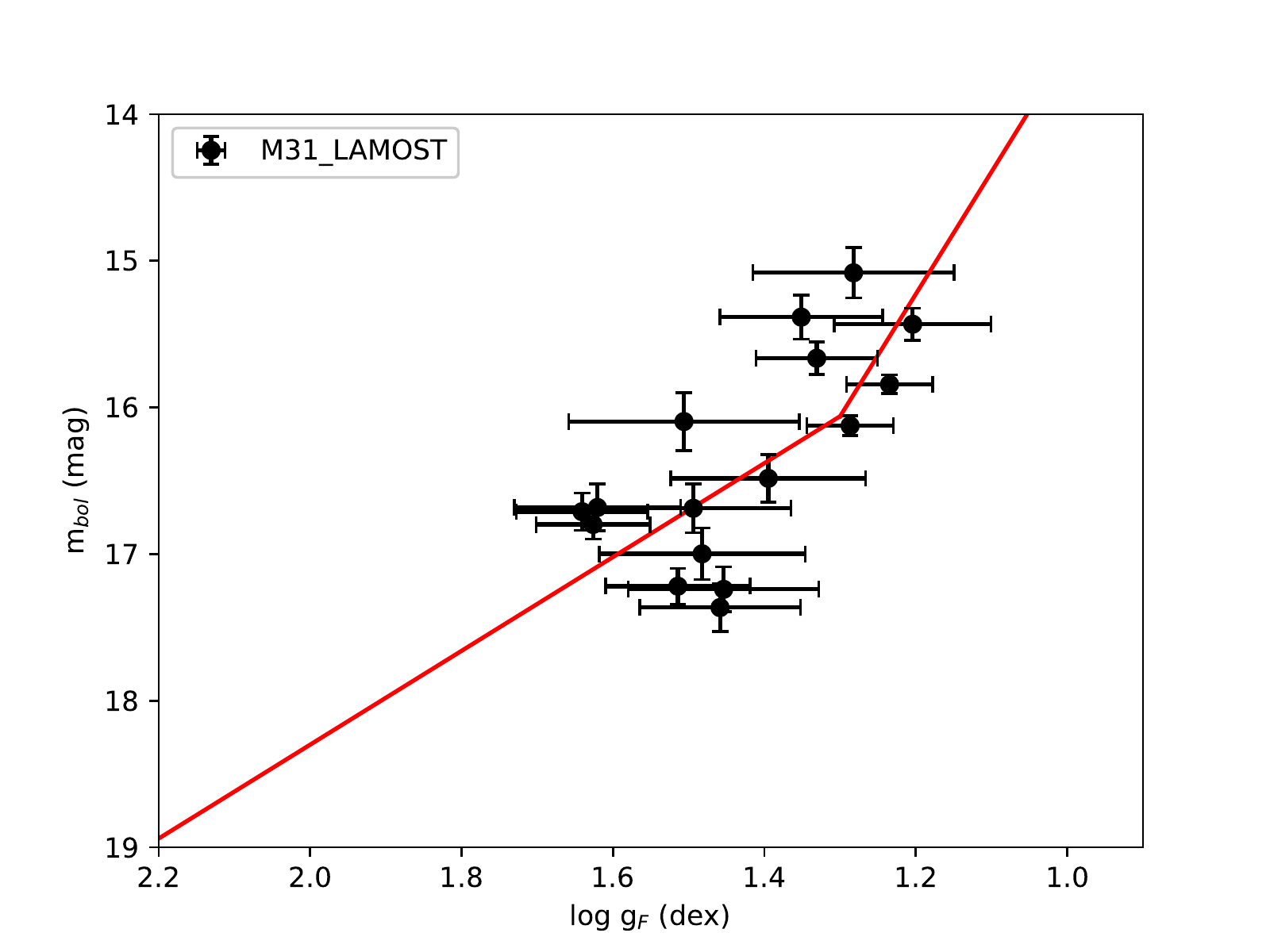}
\includegraphics[width=0.50\textwidth]{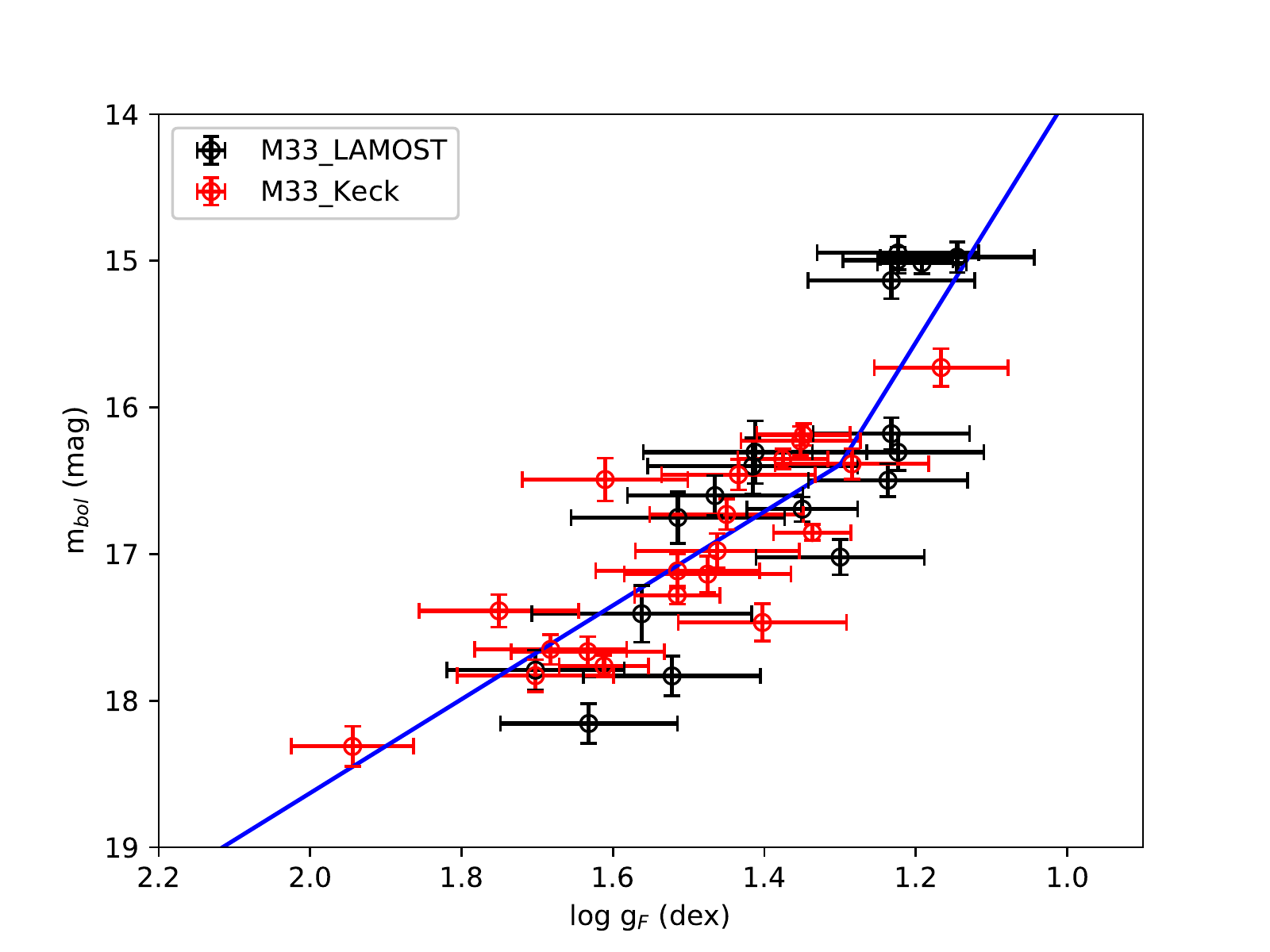}
\caption{Observed FGLR of supergiants in M 31 (top) and M 33 (bottom) with LGGS photometry. The the solid line corresponds to the LMC FGLR calibration \citep{2017AJ....154..102U} shifted to the distance moduli 24.51 mag (top) and 24.93 (bottom) mag, respectively }
\label{sect6:FGLR}
\end{figure}

Our spectroscopic analysis provides flux-weighted gravities and dereddened apparent bolometric magnitudes $m_{\rm{bol}}$ . Using the new FGLR we then determine individual distance moduli for each supergiant and obtain the final distance from a weighted mean accounting for the errors in bolometric magnitude and the logarithm of flux-weighted gravity. We obtain a distance modulus of $\mu = 24.51\pm0.13$ mag. Here 0.13 mag is the statistical error. The additional systematic errors resulting from the uncertainty of the FGLR calibration amount to 0.05 mag \cite[see][]{2016ApJ...829...70K}.
The corresponding FGLR fit is shown in Figure \ref{sect6:FGLR}. Note that we have removed the suspicious target J004005.02+403242.2 discussed above.

There are more than one hundred newly determined distance moduli published since 1990. We find that most of determined distance moduli are in the range of 24.40 to 24.60 mag consistent with our result within $1\sigma$ uncertainties. Based on a careful, statistically
weighted combination of the main stellar population tracers (Cepheids, RR Lyrae variables, and the magnitude of the TRGB), \cite{2014AJ....148...17D} derived a recommended distance modulus of  $\mu = 24.46\pm0.10$ mag by adopting a common LMC benchmark of $\mu^{\rm{LMC}} = 18.50$ mag. This recommended value agrees with the detached eclipsing binary distance modulus of $\mu = 24.36\pm0.08$ mag determined by \cite{2005ApJ...635L..37R} and \cite{2010A&A...509A..70V}.
\cite{2015MNRAS.451..724W} and \cite{2016AJ....151...88B} used near-infrared photometry photometry of Cepheids obtained by he Panchromatic Hubble Andromeda Treasury survey and derived two almost identical distance moduli of 24.51$\pm$0.08 and 24.46$\pm$0.20 mag. \cite{2016MNRAS.456..405M} applied globular cluster template fitting of two globular clusters coincident with the largest East Cloud overdensity and obtained $\mu = 24.55\pm0.05$ mag. We note that all these previous distance determinations are in agreement with our result.

\subsection{Distance to M 33}
Combining our M 33 LAMOST and Keck samples we have 38 objects to determine a distance.
We obtain a distance modulus of $\mu = 24.93\pm0.07$ mag. Figure \ref{sect6:FGLR} displays the observed M 33 FGLR and the fit with the LMC FGLR calibration.

Our result agrees well with \cite{2021ApJ...914...94S} who used the results by \cite{2009ApJ...704.1120U} combined with the Urbaneja et al. LMC calibration. This agreement is not a surprise, as both results depend on BSG FGLR fitting, except that our new sample of targets is significantly larger and the spectral analysis method is slightly different. More important is the agreement with the detached eclipsing O-star binary distance of  $\mu = 24.92\pm0.12$ mag by \cite{2006ApJ...652..313B}. We emphasize that both methods, detached eclipsing binaries and FGLR, are based on spectroscopy and, thus, not affected by large unceratinties of interstellar extinction.

We note, however, that most of the Cepheid distance determinations result in smaller values. \cite{2013ApJ...773...69G} used ground-based $J$- and $K$-band photometry in conjunction with the $V$ and $I$ photometry obtained by \cite{2001AJ....121..870M}. They derived a distance modulus $\mu = 24.62\pm0.07$ mag.  \cite{2016AJ....151...88B} applying a NIR period-Wesenheit magnitude relation obtained a very similar distance, while \cite{2011ApJS..193...26P} based on a sample of 564 Cepheids find a somewhat larger distance of $\mu = 24.76\pm0.02$ mag from a cleaned sample after quantifying biases in photometry due to crowding effects.

There are also a number of other distance determination methods in the literature. Based on a careful photometric study of the period-absolute magnitude relationship of long period variables (LPVs), \cite{2000MNRAS.313..271P} obtained a distance modulus of $\mu = 24.85\pm0.13$ mag by assuming a reddening value of $E(B-V) =0.1$ mag for M 33. Employing the TRGB method to $\it{HST}$ ACS photometry in 3 fields, \cite{2009ApJ...704.1120U} determine an average distance modulus of $\mu = 24.84\pm0.10$ mag. Other independent investigations using older stellar populations also confirmed a lager distance modulus \citep{2000AJ....120.2437S,2004ApJ...614..167C}. \cite{2000AJ....120.2437S} studied HB stars and found $\mu = 24.84\pm0.16$ mag. \cite{2004ApJ...614..167C} used the PN luminosity function to obtain $\mu = 24.86^{+0.07}_{-0.11}$ mag. 

At this point, the reason for the discrepancy between the FGLR-based and recent Cepheid distances remains an open question. We however can conclude that a large distance modulus obtained from the FGLR agrees well with other independent works based on older populations or the detached eclipsing binary.

\section{Conclusions}\label{sec:con}
We used LAMOST spectroscopic observations to study blue supergiant stars in M 31 and M 33. The targets were identified as members of this galaxy through Johnson $Q$-photometry and radial velocities after removing foreground stars and cluster. The M 33 sample was complemented by additional  supergiants observed with the Keck telescope and the DEIMOS spectrograph attached. A detailed NLTE spectral analysis was carried out to determine stellar effective temperatures, gravities and metallicities. Based on these results interstellar reddening, stellar luminosities, radii, and masses were obtained and the evolutionary status was discussed by a comparison with evolutionary tracks.

The metallicity gradients in both galaxies when scaled to the isophotal radius or the disk scale length agree well with other galaxies in the local universe and the Milky Way. However, for M 33 we find that two-dimensional distribution of metallicity deviates from azimuthal symmetry with an off-center metallicity peak located in the north-west direction.

Using a new calibration of the FGLR, we are able to determine new distances to M 31 and M 33, respectively. For M 31, we obtain an FGLR distance modulus of $\mu = 24.51\pm0.13$ mag that compares well with distances measured to Cepheids, RR Layer variables as well as bright giant stars such as TRGB or red clumps. For M 33, we obtain 24.93$\pm$0.07 mag. The FGLR-based distances are larger than distances from Cepheids studies but they agree well with other published distance moduli based on the studies of detached eclipsing binaries, LPVs, HB stars, and the PNe luminosity function.

\acknowledgments

This work was supported by the National Natural Science Foundation of China (NSFC, Nos. 11988101, 11890694, 11803048, 12033003, 11973001, 12090040, 12090044, 11903027, 11833006), National Key R\&D Program of China No. 2019YFA0405500 and Manned Space Project with NO.CMS-CSST-2021-B03 and CMS-CSST-2021-B05. C.L. and R.P.K. gratefully acknowledge support by the Munich Excellence Cluster Origins funded the Deutsche Forschungsgemeinschaft (DFG, german Research Foundation) under Germany's Excellence Strategy EXC-2094 390783311. This work was partially supported by the Scholar Program of Beijing Academy of Science and Technology (DZ:BS202002). The data presented herein were obtained at the Guoshoujing Telescope (LAMOST) and Keck Observatory. LAMOST is funded by the National Development and Reform Commission, and operated and managed by the National Astronomical Observatories, Chinese Academy of Sciences. Keck Observatory is operated as a scientific partnership among the California Institute of Technology, the University of California and the National Aeronautics and Space Administration. The Observatory was made possible by the generous financial support of the W. M. Keck Foundation. This research has made use of the NASA/IPAC Extragalactic Database (NED) and the SIMBAD Astronomical Database.

\bibliographystyle{aasjournal}
\bibliography{m31ref.bib}

%
%

\end{document}